\documentclass[10pt,a4paper,reqno]{amsart}
\usepackage{acronym}
\usepackage{amsfonts}
\usepackage{amsmath}
\usepackage{amssymb}
\usepackage{amsthm}
\usepackage{bm}
\usepackage{braket}
\usepackage{cite}
\usepackage{color}
\usepackage{geometry}
\usepackage{graphicx}
\usepackage{hyperref}
\usepackage{mathrsfs}
\usepackage{mathtools}
\usepackage{setspace}
\usepackage{stmaryrd}
\usepackage{subcaption}
\usepackage{tikz}

\usepackage{epsfig}
\usepackage{setspace}
\usepackage{booktabs}
\usepackage{threeparttable}
\usepackage{diagbox}
\usepackage{epstopdf}

\newgeometry{top=2cm,bottom=2cm,outer=1.5cm,inner=1.5cm}
\newcommand{\bse}{\begin{subequations}}
\newcommand{\ese}{\end{subequations}}

\numberwithin{equation}{section}

\title[Double and triple poles solutions for the GI type of derivative NLS equation]{Double and triple poles solutions for the Gerdjikov-Ivanov type of derivative nonlinear Schr\"{o}dinger equation with zero/nonzero boundary
conditions}
\author{Weiqi Peng}
\address[WP]{School of Mathematical Sciences, Shanghai Key Laboratory of Pure Mathematics and Mathematical Practice\\
East China Normal University \\ Shanghai 200062 \\ People's Republic of China}
\author{Yong Chen$^*$}
\address[YC]{School of Mathematical Sciences, Shanghai Key Laboratory of Pure Mathematics and Mathematical Practice\\
East China Normal University \\ Shanghai 200062 \\ People's Republic of China}
\address[YC]{College of Mathematics and Systems Science \\ Shandong University of Science and Technology \\ Qingdao 266590 \\ People's Republic of China}
\email{ychen@sei.ecnu.edu.cn(Corresponding author).}

\begin{document}

\begin{abstract}
In this work, the double and triple poles soliton solutions for the Gerdjikov-Ivanov(GI) type of derivative nonlinear Schr\"{o}dinger equation with zero boundary conditions(ZBCs) and nonzero boundary conditions(NZBCs) are studied via Riemann-Hilbert (RH) method. Though spectral problem analysis, we first give out the Jost function and scattering matrix under ZBCs and NZBCs. Then according to the analyticity, symmetry and asymptotic behavior of Jost function and scattering matrix, the Riemann-Hilbert problem(RHP) with ZBCs and NZBCs are constructed. Further, the obtained RHP with ZBCs and NZBCs can be solved in the case that reflection coefficients have double or triple poles. Finally, we derive the general precise formulae of $N$-double and $N$-triple poles solutions corresponding to ZBCs and NZBCs, respectively. The dynamical behaviors for these solutions are further discussed by image simulation.
\end{abstract}

\maketitle

\section{Introduction}
The nonlinear Schr\"{o}dinger (NLS) equation is well known to be one of the most vital integrable systems, which has been constantly reported  in various fields such as  nonlinear optics and water waves\cite{sulem,Ablowitz}. To describe more abundant physical effects and research the effects of high-order perturbations, the Gerdjikov-Ivanov (GI)  type of the derivative NLS equations was also proposed as follows\cite{GI-BJP}
\begin{align}\label{1}
iu_{t}+u_{xx}-iu^{2}u_{x}^{\ast}+\frac{1}{2}|u|^{4}u=0,
\end{align}
where the asterisk $\ast$ denotes the complex conjugation, $u$ denotes transverse magnetic field perturbation function with spatial variable $x$ and temporal variable $t$. As a extended form of the NLS equation,  the GI equation\eqref{1} is used to describe Alfve'n waves propagating parallel to the ambient magnetic field in plasma physics. The explicit soliton-like solutions for Eq.\eqref{1} are
constructed via applying its Darboux transformation\cite{Fan-GI33}, and its algebro-geometric solutions  are established  according to the Riemann theta functions\cite{Fan-GI34}. By the DT method, the breather wave and rogue wave solutions of the GI equation are
presented in Refs.\cite{Fan-GI36,He-ps}. The soliton molecules and dynamics of the smooth positons for the GI equation are discussed in Ref.\cite{Libiao-CPB}. Besides, $N$-soliton for Eq.\eqref{1} under ZBCs was given through using RH method \cite{Fan-GI38}. By using the nonlinear steepest descent method, the long-time asymptotic behavior for Eq.\eqref{1} was discussed in Refs.\cite{Guo-GI31,Guo-GI,Guo-GI34}. Recently, the RH method and Dbar-dressing method were used to construct simple pole solutions of the GI equation with NZBCs in Ref.\cite{Fan-GI39} and Ref.\cite{Fan-AML}, respectively. Higher-order pole soliton solutions of the
GI equation with ZBCs are derived by the dressing method based on
the technique of regularization\cite{Yangjianke}.  Multiple higher-order pole solutions were also obtained by the Laurent's series and generalization of the residue theorem\cite{Fan-GI40}.

In soliton theory,  the inverse scattering transform (IST)
method is the most powerful tool for analysing initial value problems of integrable nonlinear evolution equations (NLEEs). The method was raised first by Gardner et al. in 1967 for the KdV equation \cite{He-IP19}.
As is well-known, the classic IST method was based on the Gel'fand-Levitan-Marchenko(GLM) integral
equations. Subsequently, Zakharov et al. developed a RH formulation to replace the GLM equation, which simplifies the IST method\cite{Peng-ss15}. After decades of development, the RH formulation has been successfully applied to numerous integrable equations, and it is
still a hot topic today\cite{Peng-ss16,Peng-ss17,Peng-ss18,Peng-ss19,Peng-ss20,Peng-ss21,Peng-ss22,Peng-ss23,Peng-ss25,Peng-ss26,Peng-ss27}. In terms of the direct scattering, the corresponding RHP  was constructed and soliton of the can be represented by solution of the RHP. Then through solving the RHP under reflection-less case, we can obtain the formulae of multiple solitons. However, since we assume that the reflection coefficient had multiple poles, the PHP will be non-regular, which can not be solved directly by Plemelj formula. To solve this obstacle, we need to regularize the RHP via eliminating singularity. Different from the process in Refs. \cite{Yangjianke,Peng-ss19,Tian-AMP37} which transformed the non-regular RHP into the regular problem via multiplying a dressing operator, the method, pioneered by Ablowitz, Biondini, Demontis, Prinary, et al., regularized the RHP by subtracting the asymptotic behavior and the pole contribution\cite{Ablowitz,Biondini-1,Biondini-2,Biondini-3, Biondini-4, Biondini-5, Biondini-6, Biondini-7}.  Inspired by this idea, the research on the soliton solution of the integrable system  under ZBCs and NZBCs has  been complected in recent years\cite{Tian-AMP24,Tian-AMP25,Tian-AMP27,Tian-AMP28,Tian-AMP29,Tian-AMP30}. In this work, we apply this idea to derive the double and  triple poles solutions for GI equation with ZBCs and NZBCs, which is different with the simple pole solutions obtained in Refs.\cite{Fan-GI38,Fan-GI39,Fan-AML}. It is worth mentioning that the method we used is different from computing soliton solutions with higher-order poles in Refs.\cite{Yangjianke,Fan-GI40}, because  the techniques in dealing with severe spectral singularities are disparate. Moreover, for the case of NZBCs, a general double poles and triple poles solutions of GI equation with mixed discrete spectra, i.e., $2N_{1}$-breather-$2N_{2}$-soliton solutions and $3N_{1}$-breather-$3N_{2}$-soliton solutions, are analysed in this paper, which does not know yet for GI equation.  Therefore, to our knowledge, for the GI equation, the research using this idea for double and triple poles solutions under ZBCs and NZBCs has not been reported yet.

It is well known that the Lax pair of GI equation \eqref{1} can be given by
\begin{align}\label{2}
\Phi_{x}=X\Phi, \qquad \Phi_{t}=T\Phi,
\end{align}
where
\begin{align}\label{3}
&X=-ik^{2}\sigma_{3}+kQ-\frac{i}{2}Q^{2}\sigma_{3},\notag\\
&T=-2ik^{4}\sigma_{3}+2k^{3}Q-ik^{2}Q^{2}\sigma_{3}-ikQ_{x}\sigma_{3}+\frac{1}{2}(Q_{x}Q-QQ_{x})+\frac{i}{4}Q^{4}\sigma_{3},
\end{align}
and
\begin{align}\label{4}
\sigma_{3}=\left(\begin{array}{cc}
    1  &  0\\
    0 &  -1\\
\end{array}\right), \qquad Q=\left(\begin{array}{cc}
    0  &  u\\
    -u^{\ast} &  0\\
\end{array}\right),
\end{align}
where $k\in \mathbb{C}$ is a spectral parameter, and $\Phi=\Phi(x, t, k)$ is the $2\times 2$ matrix-valued
eigenfunction. Eq.\eqref{1} can be derived from the compatibility condition  $X_{t}-T_{x}+[X, T]=0$ of system \eqref{2}.

The outline of this paper is organized as follows: In section 2, we establish the RHP for the GI equation under ZBCs by spectral analysis, and the potential function is shown by the solution of the RHP. In section 3, through solving the RHP, we obtain the explicit $N$-double poles soliton solutions for the reflection coefficients with double poles under ZBCs. Additionally, the general $N$-triple poles soliton solutions for the reflection coefficients with triple poles under ZBCs are shown in section 4. In section 5, the RHP for the GI equation under NZBCs is presented via  more miscellaneous spectral problem analysis. Following that, the accurate $N$-double and $N$-triple poles solutions under NZBCs are given in section 6 and section 7, respectively.  Finally, some summaries are given in the last section.

\section{The construction of Riemann-Hilbert problem with ZBCs}
The direct scattering problem for the GI equation \eqref{1} with
ZBCs  has been discussed in Ref.\cite{Fan-GI40}. In this section, we will recall some results for the direct scattering
problem for the targeted GI equation with following ZBCs at infinity
\begin{align}\label{5}
\lim_{x\rightarrow \pm \infty}u(x, t)=0.
\end{align}

\subsection{Spectral analysis}
Let $x\rightarrow \pm \infty$, the Lax pair \eqref{2} under ZBCs \eqref{5}
changes into
\begin{align}\label{6}
\Phi_{x}=X_{0}\Phi=-ik^{2}\sigma_{3}, \qquad \Phi_{t}=T_{0}\Phi=2k^{2}X_{0}\Phi,
\end{align}
which admits the following fundamental matrix solution
\begin{align}\label{7}
\Phi^{bg}(x, t; k)=e^{-i\theta(x, t; k)\sigma_{3}},\qquad \theta(x, t; k)=k^{2}(x+2k^{2}t).
\end{align}
Defining $\Sigma:=\mathbb{R}\cup i\mathbb{R}$, the Jost solutions $\Phi_{\pm}(x, t; k)$ can be written as
\begin{align}\label{8}
\Phi_{\pm}(x, t; k)=e^{-i\theta(x, t; k)\sigma_{3}}+o(1), \quad k\in \Sigma, \quad \mbox{as} \quad x\rightarrow \pm \infty.
\end{align}
Through the variable transformation
\begin{align}\label{9}
\mu_{\pm}(x, t; k)=\Phi_{\pm}(x, t; k)e^{i\theta(x, t; k)\sigma_{3}},
\end{align}
the modified Jost solutions $\mu_{\pm}(x, t; k)$ will tend to $I$ as $x\rightarrow \pm\infty$, and it can be solved as
\begin{align}\label{10}
\mu_{\pm}(x, t; k)=I+\int_{\pm\infty}^{x}e^{-ik^{2}(x-y)\hat{\sigma}_{3}}
\left[(kQ-\frac{i}{2}Q^{2}\sigma_{3})(y, t)\mu_{\pm}(y, t; k)\right]dy.
\end{align}

\noindent \textbf{Proposition 2.1.} \emph{
Suppose $u\in L^{1} (\mathbb{R}^{\pm})$, then $\mu_{\pm}(x, t; k)$ given in Eq.\eqref{9} are unique solutions for the Jost integral equation \eqref{10}
in $\Sigma$, and they satisfy the following characteristics:}\\
\emph{$\bullet$ $\mu_{-1}(x, t; k)$ and $\mu_{+2}(x, t; k)$ become analytical for $D_{+}$ and
continuous in $D_{+}\cup \Sigma$;}\\
\emph{$\bullet$ $\mu_{+1}(x, t; k)$ and $\mu_{-2}(x, t; k)$ become analytical for $D_{-}$ and
continuous in $D_{-}\cup \Sigma$;}\\
\emph{$\bullet$ $\mu_{\pm}(x, t; k)\rightarrow I$ \mbox{as}  $k\rightarrow \infty$;}\\
\emph{$\bullet$ $\det \mu_{\pm}(x, t; k)=1, \quad x, t\in \mathbb{R}, \quad k\in \Sigma$.}

\centerline{\begin{tikzpicture}[scale=1.5]
\path [fill=gray] (-2.5,0) -- (-0.5,0) to
(-0.5,2) -- (-2.5,2);
\path [fill=gray] (-4.5,0) -- (-2.5,0) to
(-2.5,-2) -- (-4.5,-2);
\draw[-][thick](-4.5,0)--(-2.5,0);
\draw[fill] (-2.5,0) circle [radius=0.03];
\draw[->][thick](-2.5,0)--(-0.5,0)node[above]{$\mbox{Re}k$};
\draw[->][thick](-2.5,1)--(-2.5,2)node[right]{$\mbox{Im}k$};
\draw[-][thick](-2.5,1)--(-2.5,0);
\draw[-][thick](-2.5,0)--(-2.5,-1);
\draw[-][thick](-2.5,-1)--(-2.5,-2);
\draw[fill] (-2.5,-0.3) node[right]{$0$};
\draw[fill] (-1.7,0.8) circle [radius=0.03] node[right]{$k_{n}$};
\draw[fill] (-1.7,-0.8) circle [radius=0.03] node[right]{$k^{*}_{n}$};
\draw[fill] (-3.3,0.8) circle [radius=0.03] node[left]{$-k^{*}_{n}$};
\draw[fill] (-3.3,-0.8) circle [radius=0.03] node[left]{$-k_{n}$};
\end{tikzpicture}}
\noindent { \small \textbf{Figure 1.} (Color online) Distribution of the discrete spectrum and jumping curves for the RHP on complex $k$-plane, Region $D_{+}=\left\{k\in \mathbb{C} | \mbox{Re}k\mbox{Im}k> 0\right\}$ (gray region), region $D_{-}=\left\{k\in \mathbb{C} | \mbox{Re}k\mbox{Im}k< 0\right\}$ (white region).}\\

Since the Jost solutions
$\Phi_{\pm}(x, t; k)$ are the simultaneous solutions of spectral problem \eqref{2}, which satisfies following linear relation by the constant scattering matrix $S(k)=(s_{i j} (k))_{2\times 2}$
\begin{align}\label{11}
\Phi_{+}(x, t; k)=\Phi_{-}(x, t; k)S(k), \quad k\in \Sigma,
\end{align}
where $S(k)=\sigma_{2}S^{\ast}(k^{\ast})\sigma_{2}, S(k)=\sigma_{1}S^{\ast}(-k^{\ast})\sigma_{1}$, and $\sigma_{2}=\left(\begin{array}{cc}
    0  &  -i\\
  i &  0\\
\end{array}\right), \sigma_{1}=\left(\begin{array}{cc}
    0  &  1\\
  1 &  0\\
\end{array}\right).$ The scattering coefficients can be shown in what follows by Wronskians determinant
\begin{align}\label{12}
s_{11}(k)=Wr(\Phi_{+,1},\Phi_{-,2}), \quad s_{12}(k)=Wr(\Phi_{+,2},\Phi_{-,2}),\notag\\
s_{21}(k)=Wr(\Phi_{-,1},\Phi_{+,1}), \quad s_{22}(k)=Wr(\Phi_{-,1},\Phi_{+,2}),
\end{align}
where $Wr(\cdot, \cdot)$ denotes the Wronskian determinant.
From these representations, it is not hard to get following proposition.

\noindent \textbf{Proposition 2.2.} \emph{ The scattering  matrix $S(k)$ satisfies:
}\\
\emph{$\bullet$ $\det S(k)=1$ for $k\in \Sigma$;}\\
\emph{$\bullet$ $s_{22}(k)$ becomes analytical for $D_{+}$ and
continuous in  $D_{+}\cup \Sigma$;}\\
\emph{$\bullet$ $s_{11}(k)$ becomes analytical for $D_{-}$ and
continuous in  $D_{-}\cup \Sigma$;}\\
\emph{$\bullet$ $S(x,t,k)\rightarrow I$ \mbox{as}  $k\rightarrow \infty$.}

\subsection{The Riemann-Hilbert problem}
In terms of the analytic properties of Jost solutions $\mu_{\pm}(x, t; k)$
in Proposition 2.1, we have the following sectionally meromorphic matrices
\begin{align}\label{25}
M_{-}(x, t; k)=(\frac{\mu_{+,1}}{s_{11}},\mu_{-,2}),\qquad M_{+}(x, t; k)=(\mu_{-,1},\frac{\mu_{+,2}}{s_{22}}),
\end{align}
where superscripts $\pm$ represent analyticity in $D_{+}$ and $D_{-}$, respectively. Naturally, a matrix RHP is proposed:

\noindent \textbf{Riemann-Hilbert Problem 1}  \emph{
$M(x, t; k)$ solves the following RHP:
\begin{align}\label{26}
\left\{
\begin{array}{lr}
M(x, t; k)\ \mbox{is analytic in} \ \mathbb{C }\setminus \Sigma,\\
M_{-}(x, t; k)=M_{+}(x, t; k)(I-G(x, t; k)), \qquad k\in \Sigma,\\
M(x, t; k)\rightarrow I,\qquad k\rightarrow \infty,
  \end{array}
\right.
\end{align}
of which the jump matrix $G(x, t; k)$ is
\begin{align}\label{27}
G=\left(\begin{array}{cc}
    \rho(k)\tilde{\rho}(k)  &  e^{-2i\theta(x, t; k)}\tilde{\rho}(k)\\
  -e^{2i\theta(x, t; k)}\rho(k) &  0\\
\end{array}\right),
\end{align}
where $\rho(k)=\frac{s_{21}(k)}{s_{11}(k)}, \tilde{\rho}(k)=\frac{s_{12}(k)}{s_{22}(k)}$.
}

Taking
\begin{align}\label{28}
M(x, t; k)=I+\frac{1}{k}M^{(1)}(x, t; k)+O(\frac{1}{k^{2}}),\qquad k\rightarrow \infty,
\end{align}
then the potential $u(x, t)$ of the GI equation \eqref{1} with ZBCs is denoted by
\begin{align}\label{28.1}
u(x, t)=2iM_{12}^{(1)}(x, t; k)=2i\lim_{k\rightarrow\infty}k M_{12}(x, t; k).
\end{align}

\section{The solution of GI equation under ZBCs with double poles}
In this section, we will discuss the inverse scattering problem with double poles discrete spectrum for the GI equation\eqref{1} under ZBCs, and present the general $N$-double poles solutions.
\subsection{Inverse scattering problem with ZBCs and double poles}
We suppose that $s_{22}(k)$ has $N$ double zeros $k_{n}$ $(n=1, 2,\cdots, N)$ in $D_{0}=\left\{k\in\mathbb{ C}: \mbox{Re} k>0, \mbox{Im} k>0\right\}$, which means $s_{22}(k_{n})=s'_{22}(k_{n})=0$ and $s''_{22}(k_{n})\neq 0$. According to the symmetries relation of the scattering matrix, one has
\begin{align}\label{18}
\left\{
\begin{array}{lr}
s_{22}(k_{n})=s_{22}(-k_{n})=s_{11}(k_{n}^{\ast})=s_{11}(-k_{n}^{\ast})=0,\\
s'_{22}(k_{n})=s'_{22}(-k_{n})=s'_{11}(k_{n}^{\ast})=s'_{11}(-k_{n}^{\ast})=0.
\end{array}
\right.
\end{align}
Thus, the corresponding discrete spectrum can be collected as
\begin{align}\label{19}
\Gamma=\left\{k_{n}, k_{n}^{\ast}, -k_{n}^{\ast}, -k_{n}\right\}_{n=1}^{N},
\end{align}
whose distributions are displayed in Fig. 1.

Since $s_{22}(k_{0})=0$ ($k_{0}\in \Gamma\cap D_{+}$), we easily know that $\Phi_{-1}(x, t; k_{0})$ and $\Phi_{+2}(x, t; k_{0})$
are linearly dependent. Similarly, $\Phi_{+1}(x, t; k_{0})$ and $\Phi_{-2}(x, t; k_{0})$
are linearly dependent due to $s_{11}(k_{0})=0$ for $k_{0}\in \Gamma\cap D_{-}$. That is to say
\begin{align}\label{20}
\Phi_{+2}(x, t; k_{0})=b[k_{0}]\Phi_{-1}(x, t; k_{0}),\quad k_{0}\in \Gamma\cap D_{+},\notag\\
\Phi_{+1}(x, t; k_{0})=b[k_{0}]\Phi_{-2}(x, t; k_{0}),\quad k_{0}\in \Gamma\cap D_{-},
\end{align}
where $b[k_{0}]$ is a norming constant. As well as, due to $s'_{22}(k_{0})=0$ in $k_{0}\in \Gamma\cap D_{+}$, we find that $\Phi'_{+2}(x, t; k_{0})-b[k_{0}]\Phi'_{-1}(x, t; k_{0})$ and $\Phi_{-1}(x, t; k_{0})$ are linearly
dependent. Analogously,  $\Phi'_{+1}(x, t; k_{0})-b[k_{0}]\Phi'_{-2}(x, t; k_{0})$ and $\Phi_{-2}(x, t; k_{0})$ are linearly dependent for $k_{0}\in \Gamma\cap D_{-}$. Then, we obtain
\begin{align}\label{21}
\Phi'_{+2}(x, t; k_{0})-b[k_{0}]\Phi'_{-1}(x, t; k_{0})=d[k_{0}]\Phi_{-1}(x, t; k_{0}),\quad k_{0}\in \Gamma\cap D_{+},\notag\\
\Phi'_{+1}(x, t; k_{0})-b[k_{0}]\Phi'_{-2}(x, t; k_{0})=d[k_{0}]\Phi_{-2}(x, t; k_{0}),\quad k_{0}\in \Gamma\cap D_{-},
\end{align}
where $d[k_{0}]$ is also a  norming constant.  Therefore, one has
\begin{align}\label{22}
&\mathop{L_{-2}}_{k=k_{0}}\left[\frac{\Phi_{+2}(x, t; k)}{s_{22}(k)}\right]=A[k_{0}]\Phi_{-1}(x, t; k_{0}), \quad k_{0}\in \Gamma\cap D_{+},\notag\\
&\mathop{L_{-2}}_{k=k_{0}}\left[\frac{\Phi_{+1}(x, t; k)}{s_{11}(k)}\right]=A[k_{0}]\Phi_{-2}(x, t; k_{0}), \quad k_{0}\in \Gamma\cap D_{-},\notag\\
&\mathop{\mbox{Res}}_{k=k_{0}}\left[\frac{\Phi_{+2}(x, t; k)}{s_{22}(k)}\right]=A[k_{0}]\left[\Phi'_{-1}(x, t; k_{0})+B[k_{0}]\Phi_{-1}(x, t; k_{0})\right], \quad k_{0}\in \Gamma\cap D_{+},\notag\\
&\mathop{\mbox{Res}}_{k=k_{0}}\left[\frac{\Phi_{+1}(x, t; k)}{s_{11}(k)}\right]=A[k_{0}]\left[\Phi'_{-2}(x, t; k_{0})+B[k_{0}]\Phi_{-2}(x, t; k_{0})\right], \quad k_{0}\in \Gamma\cap D_{-},
\end{align}
where $L_{-2}[f(x, t; k)]$ means the coefficient of $O((k-k_{0})^{-2})$ term in the Laurent series expansion of $f(x, t; k)$ at $k=k_{0}$.
\begin{align}\label{23}
A[k_{0}]=\left\{
\begin{array}{lr}
\frac{2b[k_{0}]}{s''_{22}(k_{0})}, \quad k_{0}\in \Gamma\cap D_{+}\\
\\
\frac{2b[k_{0}]}{s''_{11}(k_{0})}, \quad k_{0}\in \Gamma\cap D_{-}.
\end{array}
\right.
\end{align}
\begin{align}\label{24}
B[k_{0}]=\left\{
\begin{array}{lr}
\frac{d[k_{0}]}{b[k_{0}]}-\frac{s'''_{22}(k_{0})}{3s''_{22}(k_{0})}, \quad k_{0}\in \Gamma\cap D_{+}\\
\\
\frac{d[k_{0}]}{b[k_{0}]}-\frac{s'''_{11}(k_{0})}{3s''_{11}(k_{0})}, \quad k_{0}\in \Gamma\cap D_{-}.
\end{array}
\right.
\end{align}

\noindent \textbf{Proposition 3.1.} \emph{ Let $k_{0}\in \Gamma$, then
the following symmetry relations are satisfied:
}\\
\emph{$\bullet$ The first symmetry relation $A[k_{0}]=-A[k^{\ast}_{0}]^{\ast}, B[k_{0}]=B[k^{\ast}_{0}]^{\ast}$;}\\
\emph{$\bullet$ The second symmetry relation $A[k_{0}]=A[-k^{\ast}_{0}]^{\ast}, B[k_{0}]=-B[-k^{\ast}_{0}]^{\ast}$.}

In order to solve the RHP conveniently, we take
\begin{align}\label{29}
\xi_{n}=\left\{
\begin{array}{lr}
k_{n}, \qquad n=1, 2, \cdots N\\
-k_{n-N}, \qquad n=N+1, N+2, \cdots 2N.
\end{array}
\right.
\end{align}
Then the residue and the coefficient $L_{-2}$ of $M(x, t; k)$ can be expressed as

\begin{align}\label{31}
&\mathop{L_{-2}}_{k=\xi_{n}}M_{+}=\left(0, A[\xi_{n}]e^{-2i\theta(x, t; \xi_{n})}\mu_{-1}(x, t; \xi_{n})\right),\notag\\
&\mathop{L_{-2}}_{k=\xi_{n}^{\ast}}M_{-}=\left(A[\xi_{n}^{\ast}]e^{2i\theta(x, t; \xi_{n}^{\ast})}\mu_{-2}(x, t; \xi_{n}^{\ast}), 0\right),\notag\\
&\mathop{\mbox{Res}}_{k=\xi_{n}}M_{+}=\left(0, A[\xi_{n}]e^{-2i\theta(x, t; \xi_{n})}\left[\mu'_{-1}(x, t; \xi_{n})+\left[B[\xi_{n}]-2i\theta'(x, t; \xi_{n})\right]\mu_{-1}(x, t; \xi_{n})\right]\right),\notag\\
&\mathop{\mbox{Res}}_{k=\xi_{n}^{\ast}}M_{-}=\left(A[\xi_{n}^{\ast}]e^{2i\theta(x, t; \xi_{n}^{\ast})}\left[\mu'_{-2}(x, t; \xi_{n}^{\ast})+\left[B[\xi_{n}^{\ast}]+2i\theta'(x, t; \xi_{n}^{\ast})\right]\mu_{-2}(x, t; \xi_{n}^{\ast})\right], 0\right).
\end{align}

By subtracting out the residue, the coefficient $L_{-2}$ and the asymptotic values as $k\rightarrow\infty$ from the original non-regular RHP, one can obtain the following regular RHP
\begin{gather}
M_{-}-I-\sum_{n=1}^{2N}\left[\frac{\mathop{L_{-2}}\limits_{k=\xi_{n}}M_{+}}{(k-\xi_{n})^{2}}
+\frac{\mathop{\mbox{Res}}\limits_{k=\xi_{n}}M_{+}}{k-\xi_{n}}+\frac{\mathop{L_{-2}}\limits_{k=\xi_{n}^{\ast}}M_{-}}
{(k-\xi_{n}^{\ast})^{2}}+\frac{\mathop{\mbox{Res}}\limits_{k=\xi_{n}^{\ast}}M_{-}}{(k-\xi_{n}^{\ast})}\right]=\notag\\
M_{+}-I-\sum_{n=1}^{2N}\left[\frac{\mathop{L_{-2}}\limits_{k=\xi_{n}}M_{+}}{(k-\xi_{n})^{2}}
+\frac{\mathop{\mbox{Res}}\limits_{k=\xi_{n}}M_{+}}{k-\xi_{n}}+\frac{\mathop{L_{-2}}\limits_{k=\xi_{n}^{\ast}}M_{-}}
{(k-\xi_{n}^{\ast})^{2}}+\frac{\mathop{\mbox{Res}}\limits_{k=\xi_{n}^{\ast}}M_{-}}{(k-\xi_{n}^{\ast})}\right]-M_{+}G,\label{32}
\end{gather}
which can be solved by the Plemelj's formulae, given by
\begin{align}\label{33}
M(x, t; k)=&I+\sum_{n=1}^{2N}\left[\frac{\mathop{L_{-2}}\limits_{k=\xi_{n}}M_{+}}{(k-\xi_{n})^{2}}
+\frac{\mathop{\mbox{Res}}\limits_{k=\xi_{n}}M_{+}}{k-\xi_{n}}+\frac{\mathop{L_{-2}}\limits_{k=\xi_{n}^{\ast}}M_{-}}
{(k-\xi_{n}^{\ast})^{2}}+\frac{\mathop{\mbox{Res}}\limits_{k=\xi_{n}^{\ast}}M_{-}}{(k-\xi_{n}^{\ast})}\right]\notag\\
&+\frac{1}{2\pi i}\int_{\Sigma}\frac{M_{+}(x, t; \zeta)G(x, t; \zeta)}{\zeta-k}d\zeta,\qquad k\in \mathbb{C}\backslash \Sigma,
\end{align}
where
\begin{gather}
\frac{\mathop{L_{-2}}\limits_{k=\xi_{n}}M_{+}}{(k-\xi_{n})^{2}}
+\frac{\mathop{\mbox{Res}}\limits_{k=\xi_{n}}M_{+}}{k-\xi_{n}}+\frac{\mathop{L_{-2}}\limits_{k=\xi_{n}^{\ast}}M_{-}}
{(k-\xi_{n}^{\ast})^{2}}+\frac{\mathop{\mbox{Res}}\limits_{k=\xi_{n}^{\ast}}M_{-}}{(k-\xi_{n}^{\ast})}=\notag\\
\left(\hat{C}_{n}(k)\left[\mu'_{-2}(\xi_{n}^{\ast})+\left(\hat{D}_{n}+\frac{1}{k-\xi_{n}^{\ast}}\right)\mu_{-2}(\xi_{n}^{\ast})\right],
C_{n}(k)\left[\mu'_{-1}(\xi_{n})+\left(D_{n}+\frac{1}{k-\xi_{n}}\right)\mu_{-1}(\xi_{n})\right]\right),\label{34}
\end{gather}
and
\begin{align}\label{35}
&C_{n}(k)=\frac{A[\xi_{n}]e^{-2i\theta(\xi_{n})}}{k-\xi_{n}}, \quad D_{n}=B[\xi_{n}]-2i\theta'(\xi_{n}), \notag\\
&\hat{C}_{n}(k)=\frac{A[\xi_{n}^{\ast}]e^{2i\theta(\xi_{n}^{\ast})}}{k-\xi_{n}^{\ast}}, \quad \hat{D}_{n}=B[\xi_{n}^{\ast}]+2i\theta'(\xi_{n}^{\ast}).
\end{align}
Furthermore, according to \eqref{28}, one has
\begin{gather}
M^{(1)}(x, t; k)=-\frac{1}{2\pi i}\int_{\Sigma}M_{+}(x, t; \zeta)G(x, t; \zeta)d\zeta+ \notag\\
\sum_{n=1}^{2N}\left(A[\xi_{n}^{\ast}]e^{2i\theta(\xi_{n}^{\ast})}\left(\mu'_{-2}(\xi_{n}^{\ast})+\hat{D}_{n}\mu_{-2}(\xi_{n}^{\ast})\right),
A[\xi_{n}]e^{-2i\theta(\xi_{n})}\left(\mu'_{-1}(\xi_{n})+D_{n}\mu_{-1}(\xi_{n})\right)\right). \label{37}
\end{gather}
The potential $u(x, t)$ with double poles for the GI equation with ZBCs is redefined into
\begin{gather}
u(x, t)=2iM_{12}^{(1)}=-\frac{1}{\pi}\int_{\Sigma}(M_{+}(x, t; \zeta)G(x, t; \zeta))_{12}d\zeta \notag\\
+2i\sum_{n=1}^{2N}A[\xi_{n}]e^{-2i\theta(\xi_{n})}\left(\mu'_{-11}(\xi_{n})+D_{n}\mu_{-11}(\xi_{n})\right)
.\label{38}
\end{gather}

\subsection{Double poles soliton solutions with ZBCs}
To derive the explicit double poles soliton solutions of the GI equation with ZBCs, we take $\rho(k)=\tilde{\rho}(k)=0$ called the reflectionless. Then, the second column of Eq.\eqref{33} yields
\begin{align}\label{39}
&\mu_{-2}(\xi_{j}^{\ast})=\left(\begin{array}{c}
     0\\
  1\\
\end{array}\right)+\sum_{n=1}^{2N}C_{n}(\xi_{j}^{\ast})\left[\mu'_{-1}(\xi_{n})+\left(D_{n}+\frac{1}{\xi_{j}^{\ast}-\xi_{n}}\right)\mu_{-1}(\xi_{n})\right],
\notag\\
&\mu_{-2}'(\xi_{j}^{\ast})=-\sum_{n=1}^{2N}\frac{C_{n}(\xi_{j}^{\ast})}{\xi_{j}^{\ast}-\xi_{n}}\left[\mu'_{-1}(\xi_{n})+\left(D_{n}+\frac{2}{\xi_{j}^{\ast}-\xi_{n}}\right)\mu_{-1}(\xi_{n})\right],
\notag\\
&\mu_{-1}(\xi_{j})=\left(\begin{array}{c}
     1\\
  0\\
\end{array}\right)+\sum_{n=1}^{2N}\hat{C}_{n}(\xi_{j})\left[\mu'_{-2}(\xi_{n}^{\ast})+\left(\hat{D}_{n}+\frac{1}{\xi_{j}-\xi_{n}^{\ast}}\right)\mu_{-2}(\xi_{n}^{\ast})\right],
\notag\\
&\mu_{-1}'(\xi_{j})=-\sum_{n=1}^{2N}\frac{\hat{C}_{n}(\xi_{j})}{\xi_{j}-\xi_{n}^{\ast}}\left[\mu'_{-2}(\xi_{n}^{\ast})+\left(\hat{D}_{n}+\frac{2}{\xi_{j}-\xi_{n}^{\ast}}\right)\mu_{-2}(\xi_{n}^{\ast})\right].
\end{align}
Then, we have
\begin{align}\label{40}
&\mu_{-12}(\xi_{j}^{\ast})=\sum_{n=1}^{2N}C_{n}(\xi_{j}^{\ast})\left[\mu'_{-11}(\xi_{n})+\left(D_{n}+\frac{1}{\xi_{j}^{\ast}-\xi_{n}}\right)\mu_{-11}(\xi_{n})\right],
\notag\\
&\mu_{-12}'(\xi_{j}^{\ast})=-\sum_{n=1}^{2N}\frac{C_{n}(\xi_{j}^{\ast})}{\xi_{j}^{\ast}-\xi_{n}}\left[\mu'_{-11}(\xi_{n})+\left(D_{n}+\frac{2}{\xi_{j}^{\ast}-\xi_{n}}\right)\mu_{-11}(\xi_{n})\right],
\notag\\
&\mu_{-11}(\xi_{j})=1+\sum_{n=1}^{2N}\hat{C}_{n}(\xi_{j})\left[\mu'_{-12}(\xi_{n}^{\ast})+\left(\hat{D}_{n}+\frac{1}{\xi_{j}-\xi_{n}^{\ast}}\right)\mu_{-12}(\xi_{n}^{\ast})\right],
\notag\\
&\mu_{-11}'(\xi_{j})=-\sum_{n=1}^{2N}\frac{\hat{C}_{n}(\xi_{j})}{\xi_{j}-\xi_{n}^{\ast}}\left[\mu'_{-12}(\xi_{n}^{\ast})+\left(\hat{D}_{n}+\frac{2}{\xi_{j}-\xi_{n}^{\ast}}\right)\mu_{-12}(\xi_{n}^{\ast})\right],
\end{align}
\noindent \textbf{Theorem 3.1}  \emph{
The explicit formula of the double poles soliton solution for the GI equation \eqref{1} with ZBCs \eqref{5} is expressed as
\begin{align}\label{41}
u(x,t)=-2i\frac{\det \left(\begin{array}{cc}
    I-H  &  \beta\\
  \alpha^{T} &  0\\
\end{array}\right)}{\det (I-H)},
\end{align}
where $\beta, H, \alpha$ are given in \eqref{43.1}, \eqref{42} and \eqref{44.1}.
}
\begin{proof}
We can rewrite the  linear system  \eqref{40}  in the matrix form:
\begin{align}\label{43}
\psi-H\psi=\beta,
\end{align}
where
\begin{gather}
\psi=\left(\begin{array}{c}
     \psi^{(1)}\\
  \psi^{(2)}\\
\end{array}\right), \psi^{(1)}=(\mu_{-11}(\xi_{1}),\cdots, \mu_{-11}(\xi_{2N}))^{T},  \psi^{(2)}=(\mu'_{-11}(\xi_{1}),\cdots, \mu'_{-11}(\xi_{2N}))^{T},\notag\\
 \beta=\left(\begin{array}{c}
     \beta^{(1)}\\
  \beta^{(2)}\\
\end{array}\right), \beta^{(1)}=(1)_{2N\times 1}, \beta^{(2)}=(0)_{2N\times 1}. \label{43.1}
\end{gather}
The $4N\times 4N$ matrix $H=\left(\begin{array}{cc}
    H^{(11)} &  H^{(12)}\\
  H^{(21)} &  H^{(22)}\\
\end{array}\right)$ with $H^{(im)}=\left(H^{(im)}_{jk}\right)_{2N\times 2N}(i, m=
1, 2)$ given by
\begin{align}\label{42}
&H^{(11)}_{jk}=\sum_{n=1}^{2N}\hat{C}_{n}(\xi_{j})C_{k}(\xi_{n}^{\ast})\left[-\frac{1}{\xi_{n}^{\ast}-\xi_{k}}(D_{k}+\frac{2}{\xi_{n}^{\ast}-\xi_{k}})
+(\hat{D}_{n}+\frac{1}{\xi_{j}-\xi_{n}^{\ast}})(D_{k}+\frac{1}{\xi_{n}^{\ast}-\xi_{k}})\right]\notag\\
&H^{(12)}_{jk}=\sum_{n=1}^{2N}\hat{C}_{n}(\xi_{j})C_{k}(\xi_{n}^{\ast})\left[-\frac{1}{\xi_{n}^{\ast}-\xi_{k}}+(\hat{D}_{n}+\frac{1}{\xi_{j}-\xi_{n}^{\ast}})\right]\notag\\
&H^{(21)}_{jk}=\sum_{n=1}^{2N}\frac{\hat{C}_{n}(\xi_{j})C_{k}(\xi_{n}^{\ast})}{\xi_{j}-\xi_{n}^{\ast}}\left[\frac{1}{\xi_{n}^{\ast}-\xi_{k}}(D_{k}+\frac{2}{\xi_{n}^{\ast}-\xi_{k}})
-(\hat{D}_{n}+\frac{2}{\xi_{j}-\xi_{n}^{\ast}})(D_{k}+\frac{1}{\xi_{n}^{\ast}-\xi_{k}})\right]\notag\\
&H^{(22)}_{jk}=\sum_{n=1}^{2N}\frac{\hat{C}_{n}(\xi_{j})C_{k}(\xi_{n}^{\ast})}{\xi_{j}-\xi_{n}^{\ast}}\left[\frac{1}{\xi_{n}^{\ast}-\xi_{k}}-(\hat{D}_{n}+\frac{2}{\xi_{j}-\xi_{n}^{\ast}})\right].
\end{align}
According to the reflectionless potential, Eq.\eqref{38} can be rewritten as
\begin{align}\label{44}
u=2i\alpha^{T} \psi,
\end{align}
where
\begin{gather}
\alpha=\left(\begin{array}{c}
     \alpha^{(1)}\\
  \alpha^{(2)}\\
\end{array}\right),   \alpha^{(2)}=(A[\xi_{1}]e^{-2i\theta(\xi_{1})}, A[\xi_{2}]e^{-2i\theta(\xi_{2})},\cdots, A[\xi_{2N}]e^{-2i\theta(\xi_{2N})})^{T},\notag\\
\alpha^{(1)}=(A[\xi_{1}]e^{-2i\theta(\xi_{1})}D_{1}, A[\xi_{2}]e^{-2i\theta(\xi_{2})}D_{2},\cdots, A[\xi_{2N}]e^{-2i\theta(\xi_{2N})}D_{2N})^{T}.\label{44.1}
\end{gather}
Combining Eqs. \eqref{43}, the  expression  of the double poles soliton solution can be derived.
\end{proof}

We will analyse the dynamical behaviors of the double poles soliton solution for GI equation with ZBCs. At the case of $N = 1$,  we display the following figures by choosing suitable parameters. From Fig.2 (a)(b), we easily find that the one-double poles soliton solution is actually a kind of  bound-state soliton solution which represents the interaction of two bright soliton. Besides, the interaction between two bright soliton is  an elastic collision with  the shape and size of the soliton unchange. Fig. 2(c) presents the wave propagation along the $x$-axis at $t = -5, 0, 5$. Fig.3 displays  the interaction of two pairs of one-double soliton solutions,i.e., two-double soliton solutions, at the case of $N=2$.\\
{\rotatebox{0}{\includegraphics[width=5.0cm,height=5.0cm,angle=0]{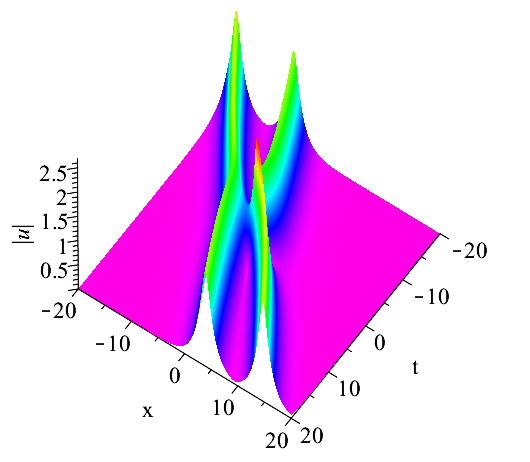}}}
~~~~\quad\qquad
{\rotatebox{0}{\includegraphics[width=5.0cm,height=5.0cm,angle=0]{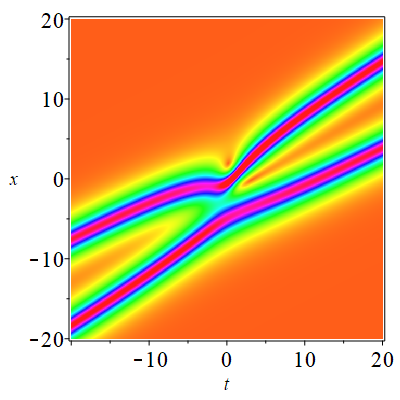}}}
~~~~\quad\qquad
{\rotatebox{0}{\includegraphics[width=5.0cm,height=5.0cm,angle=0]{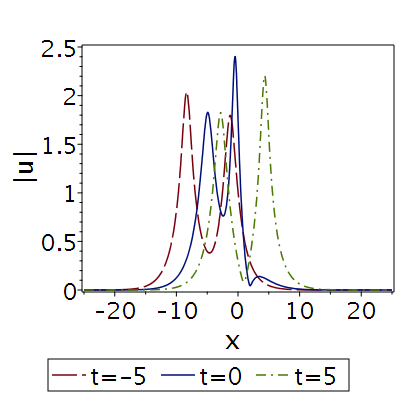}}}

$\qquad\qquad\textbf{(a)}\qquad\qquad\qquad\qquad\qquad\qquad\quad\qquad\qquad\quad 
\textbf{(b)}\qquad\qquad\qquad\qquad\qquad\qquad\qquad\qquad\quad\textbf{(c)}$\\
\noindent { \small \textbf{Figure 2.} (Color online) The one-double pole soliton solution for Eq.\eqref{1} with ZBCs and $N=1$. The parameters are $A[k_{1}]=1, B[k_{1}]=i, k_{1}=\frac{1}{3}+\frac{1}{2}i$. $\textbf{(a)}$ Three dimensional plot;
$\textbf{(b)}$ The density plot;
$\textbf{(c)}$ The wave propagation along the $x$-axis at $t=-5$(longdash), $t=0$(solid), $t=5$(dashdot).}\\
{\rotatebox{0}{\includegraphics[width=5.0cm,height=5.0cm,angle=0]{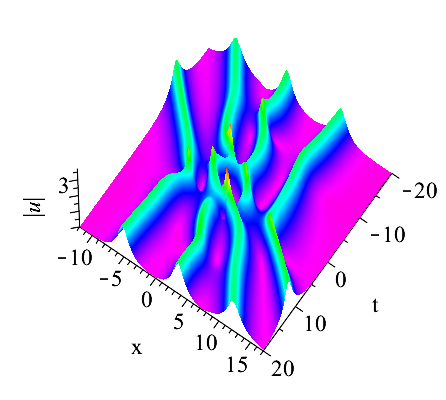}}}
~~~~\quad\qquad
{\rotatebox{0}{\includegraphics[width=5.0cm,height=5.0cm,angle=0]{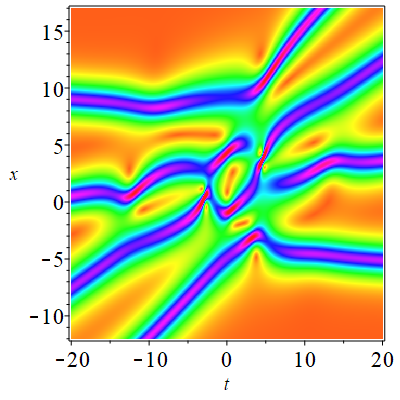}}}
~~~~\quad\qquad
{\rotatebox{0}{\includegraphics[width=5.0cm,height=5.0cm,angle=0]{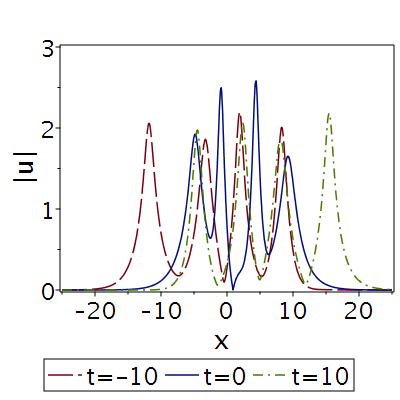}}}

$\qquad\qquad\textbf{(a)}\qquad\qquad\qquad\qquad\qquad\qquad\quad\qquad\qquad\quad 
\textbf{(b)}\qquad\qquad\qquad\qquad\qquad\qquad\qquad\qquad\quad\textbf{(c)}$\\
\noindent { \small \textbf{Figure 3.} (Color online) The two-double pole soliton solution for Eq.\eqref{1} with ZBCs and $N=2$. The parameters are $A[k_{1}]=1, B[k_{1}]=1, A[k_{2}]=1, B[k_{2}]=1, k_{1}=\frac{1}{2}+\frac{1}{2}i, k_{2}=\frac{1}{3}+\frac{1}{2}i$. $\textbf{(a)}$ Three dimensional plot;
$\textbf{(b)}$ The density plot;
$\textbf{(c)}$ The wave propagation along the $x$-axis at $t=-10$(longdash), $t=0$(solid), $t=10$(dashdot).}\\

\section{The solution of GI equation under ZBCs with triple poles}
In this section, we devote to derive the general $N$-triple poles solutions through  discussing the inverse scattering problem with triple poles discrete spectrum for the GI equation\eqref{1} under ZBCs.
\subsection{Inverse scattering problem with ZBCs and triple poles}
The triple zeros $k_{n}$ $(n=1, 2,\cdots, N)$ in $D_{0}=\left\{k\in\mathbb{ C}: \mbox{Re} k>0, \mbox{Im} k>0\right\}$ mean $s_{22}(k_{n})=s'_{22}(k_{n})=s''_{22}(k_{n})=0$ and $s'''_{22}(k_{n})\neq 0$. Be similar to the expressions \eqref{20}, \eqref{21}, and due to $s''_{22}(k_{0})=0$ in $k_{0}\in \Gamma\cap D_{+}$, the $\Phi''_{+2}(x, t; k_{0})-b[k_{0}]\Phi''_{-1}(x, t; k_{0})-2d[k_{0}]\Phi'_{-1}(x, t; k_{0})$ and $\Phi_{-1}(x, t; k_{0})$ are linearly
dependent. Besides,  $\Phi''_{+1}(x, t; k_{0})-b[k_{0}]\Phi''_{-2}(x, t; k_{0})-2d[k_{0}]\Phi'_{-2}(x, t; k_{0})$ and $\Phi_{-2}(x, t; k_{0})$ are linearly dependent for $k_{0}\in \Gamma\cap D_{-}$. Naturally, we get
\begin{align}\label{21a}
\Phi''_{+2}(x, t; k_{0})-b[k_{0}]\Phi''_{-1}(x, t; k_{0})-2d[k_{0}]\Phi'_{-1}(x, t; k_{0})=h[k_{0}]\Phi_{-1}(x, t; k_{0}),\quad k_{0}\in \Gamma\cap D_{+},\notag\\
\Phi''_{+1}(x, t; k_{0})-b[k_{0}]\Phi''_{-2}(x, t; k_{0})-2d[k_{0}]\Phi'_{-2}(x, t; k_{0})=h[k_{0}]\Phi_{-2}(x, t; k_{0}),\quad k_{0}\in \Gamma\cap D_{-},
\end{align}
where $h[k_{0}]$ is also a norming constant.  Thus, one has
\begin{align}\label{22a}
&\mathop{L_{-3}}_{k=k_{0}}\left[\frac{\Phi_{+2}(x, t; k)}{s_{22}(k)}\right]=\tilde{A}[k_{0}]\Phi_{-1}(x, t; k_{0}), \quad k_{0}\in \Gamma\cap D_{+},\notag\\
&\mathop{L_{-3}}_{k=k_{0}}\left[\frac{\Phi_{+1}(x, t; k)}{s_{11}(k)}\right]=\tilde{A}[k_{0}]\Phi_{-2}(x, t; k_{0}), \quad k_{0}\in \Gamma\cap D_{-},\notag\\
&\mathop{L_{-2}}_{k=k_{0}}\left[\frac{\Phi_{+2}(x, t; k)}{s_{22}(k)}\right]=\tilde{A}[k_{0}]\left[\Phi'_{-1}(x, t; k_{0})+\tilde{B}[k_{0}]\Phi_{-1}(x, t; k_{0})\right], \quad k_{0}\in \Gamma\cap D_{+},\notag\\
&\mathop{L_{-2}}_{k=k_{0}}\left[\frac{\Phi_{+1}(x, t; k)}{s_{11}(k)}\right]=\tilde{A}[k_{0}]\left[\Phi'_{-2}(x, t; k_{0})+\tilde{B}[k_{0}]\Phi_{-2}(x, t; k_{0})\right], \quad k_{0}\in \Gamma\cap D_{-},\notag\\
&\mathop{\mbox{Res}}_{k=k_{0}}\left[\frac{\Phi_{+2}(x, t; k)}{s_{22}(k)}\right]=\tilde{A}[k_{0}]\left[\frac{1}{2}\Phi''_{-1}(x, t; k_{0})+\tilde{B}[k_{0}]\Phi'_{-1}(x, t; k_{0})+\tilde{C}[k_{0}]\Phi_{-1}(x, t; k_{0})\right], \quad k_{0}\in \Gamma\cap D_{+},\notag\\
&\mathop{\mbox{Res}}_{k=k_{0}}\left[\frac{\Phi_{+1}(x, t; k)}{s_{11}(k)}\right]=\tilde{A}[k_{0}]\left[\frac{1}{2}\Phi''_{-2}(x, t; k_{0})+\tilde{B}[k_{0}]\Phi'_{-2}(x, t; k_{0})+\tilde{C}[k_{0}]\Phi_{-2}(x, t; k_{0})\right], \quad k_{0}\in \Gamma\cap D_{-},
\end{align}
where $L_{-3}[f(x, t; k)]$ denotes the coefficient of $O((k-k_{0})^{-3})$ term in the Laurent series expansion of $f(x, t; k)$ at $k=k_{0}$.
\begin{align}\label{23a}
\tilde{A}[k_{0}]=\left\{
\begin{array}{lr}
\frac{6b[k_{0}]}{s'''_{22}(k_{0})}, \quad k_{0}\in \Gamma\cap D_{+}\\
\\
\frac{6b[k_{0}]}{s'''_{11}(k_{0})}, \quad k_{0}\in \Gamma\cap D_{-}.
\end{array}
\right.
\end{align}
\begin{align}\label{24a}
\tilde{B}[k_{0}]=\left\{
\begin{array}{lr}
\frac{d[k_{0}]}{b[k_{0}]}-\frac{s''''_{22}(k_{0})}{4s'''_{22}(k_{0})}, \quad k_{0}\in \Gamma\cap D_{+}\\
\\
\frac{d[k_{0}]}{b[k_{0}]}-\frac{s''''_{11}(k_{0})}{4s'''_{11}(k_{0})}, \quad k_{0}\in \Gamma\cap D_{-}.
\end{array}
\right.
\end{align}
\begin{align}\label{24a1}
\tilde{C}[k_{0}]=\left\{
\begin{array}{lr}
\frac{h[k_{0}]}{2b[k_{0}]}-\frac{d[k_{0}]s''''_{22}(k_{0})}{4b[k_{0}]s'''_{22}(k_{0})}+\frac{(s''''_{22})^{2}(k_{0})}{16(s'''_{22})^{2}(k_{0})}, \quad k_{0}\in \Gamma\cap D_{+}\\
\\
\frac{h[k_{0}]}{2b[k_{0}]}-\frac{d[k_{0}]s''''_{11}(k_{0})}{4b[k_{0}]s'''_{11}(k_{0})}+\frac{(s''''_{11})^{2}(k_{0})}{16(s'''_{11})^{2}(k_{0})}, \quad k_{0}\in \Gamma\cap D_{-}.
\end{array}
\right.
\end{align}

\noindent \textbf{Proposition 4.1.} \emph{ Let $k_{0}\in \Gamma$, then
the following symmetry relations are satisfied:
}\\
\emph{$\bullet$ The first symmetry relation $\tilde{A}[k_{0}]=-\tilde{A}[k^{\ast}_{0}]^{\ast}, \tilde{B}[k_{0}]=\tilde{B}[k^{\ast}_{0}]^{\ast}, \tilde{C}[k_{0}]=\tilde{C}[k^{\ast}_{0}]^{\ast}$;}\\
\emph{$\bullet$ The second symmetry relation $\tilde{A}[k_{0}]=-\tilde{A}[-k^{\ast}_{0}]^{\ast}, \tilde{B}[k_{0}]=-\tilde{B}[-k^{\ast}_{0}]^{\ast}, \tilde{C}[k_{0}]=\tilde{C}[-k^{\ast}_{0}]^{\ast}$.}

Then the residue, the coefficient $L_{-2}$  and the coefficient $L_{-3}$ of $M(x, t; k)$ can be written as
\begin{align}\label{31a}
&\mathop{L_{-3}}_{k=\xi_{n}}M_{+}=\left(0, \tilde{A}[\xi_{n}]e^{-2i\theta(\xi_{n})}\mu_{-1}(\xi_{n})\right),\notag\\
&\mathop{L_{-3}}_{k=\xi_{n}^{\ast}}M_{-}=\left(\tilde{A}[\xi_{n}^{\ast}]e^{2i\theta(\xi_{n}^{\ast})}\mu_{-2}(\xi_{n}^{\ast}), 0\right),\notag\\
&\mathop{L_{-2}}_{k=\xi_{n}}M_{+}=\left(0, \tilde{A}[\xi_{n}]e^{-2i\theta(\xi_{n})}\left[\mu'_{-1}(x, t; \xi_{n})+\left[\tilde{B}[\xi_{n}]-2i\theta'(\xi_{n})\right]\mu_{-1}(\xi_{n})\right]\right),\notag\\
&\mathop{L_{-2}}_{k=\xi_{n}^{\ast}}M_{-}=\left(\tilde{A}[\xi_{n}^{\ast}]e^{2i\theta(\xi_{n}^{\ast})}\left[\mu'_{-2}( \xi_{n}^{\ast})+\left[\tilde{B}[\xi_{n}^{\ast}]+2i\theta'(\xi_{n}^{\ast})\right]\mu_{-2}(\xi_{n}^{\ast})\right], 0\right),\notag\\
&\mathop{\mbox{Res}}_{k=\xi_{n}}M_{+}=\left(0, \tilde{A}[\xi_{n}]e^{-2i\theta(\xi_{n})}\left[\frac{1}{2}\mu''_{-1}( \xi_{n})+\left[\tilde{B}[\xi_{n}]-2i\theta'(\xi_{n})\right]\mu'_{-1}(\xi_{n})+[\tilde{C}[\xi_{n}]-\Theta_{1}(\xi_{n})]\mu_{-1}( \xi_{n})\right]\right),\notag\\
&\mathop{\mbox{Res}}_{k=\xi_{n}^{\ast}}M_{-}=\left(\tilde{A}[\xi_{n}^{\ast}]e^{2i\theta(\xi_{n}^{\ast})}\left[\frac{1}{2}\mu''_{-2}( \xi^{\ast}_{n})+\left[\tilde{B}[\xi^{\ast}_{n}]+2i\theta'(\xi^{\ast}_{n})\right]\mu'_{-2}( \xi^{\ast}_{n})+[\tilde{C}[\xi^{\ast}_{n}]-\Theta_{2}(\xi^{\ast}_{n})]\mu_{-2}(\xi^{\ast}_{n})\right], 0\right),
\end{align}
where
\begin{align}\label{31a1}
&\Theta_{1}(\xi_{n})=2(\theta'(\xi_{n}))^{2}+i\theta''(\xi_{n})+2\tilde{B}[\xi_{n}]i\theta'(\xi_{n}),\notag\\
&\Theta_{2}(\xi^{\ast}_{n})=2(\theta'(\xi^{\ast}_{n}))^{2}-i\theta''(\xi^{\ast}_{n})-2\tilde{B}[\xi^{\ast}_{n}]i\theta'(\xi^{\ast}_{n}).
\end{align}
Subtracting out the residue, the coefficient $L_{-2}, L_{-3}$ and the asymptotic values as $k\rightarrow\infty$ from the original non-regular RHP, the regular RHP is derived
\begin{gather}
M_{-}-I-\notag\\
\sum_{n=1}^{2N}\left[\frac{\mathop{L_{-3}}\limits_{k=\xi_{n}}M_{+}}{(k-\xi_{n})^{3}}
+\frac{\mathop{L_{-2}}\limits_{k=\xi_{n}}M_{+}}{(k-\xi_{n})^{2}}
+\frac{\mathop{\mbox{Res}}\limits_{k=\xi_{n}}M_{+}}{k-\xi_{n}}+\frac{\mathop{L_{-3}}\limits_{k=\xi_{n}^{\ast}}M_{-}}
{(k-\xi_{n}^{\ast})^{3}}+\frac{\mathop{L_{-2}}\limits_{k=\xi_{n}^{\ast}}M_{-}}
{(k-\xi_{n}^{\ast})^{2}}+\frac{\mathop{\mbox{Res}}\limits_{k=\xi_{n}^{\ast}}M_{-}}{(k-\xi_{n}^{\ast})}\right]=\notag\\
M_{+}-I-\notag\\
\sum_{n=1}^{2N}\left[\frac{\mathop{L_{-3}}\limits_{k=\xi_{n}}M_{+}}{(k-\xi_{n})^{3}}
+\frac{\mathop{L_{-2}}\limits_{k=\xi_{n}}M_{+}}{(k-\xi_{n})^{2}}
+\frac{\mathop{\mbox{Res}}\limits_{k=\xi_{n}}M_{+}}{k-\xi_{n}}+\frac{\mathop{L_{-3}}\limits_{k=\xi_{n}^{\ast}}M_{-}}
{(k-\xi_{n}^{\ast})^{3}}+\frac{\mathop{L_{-2}}\limits_{k=\xi_{n}^{\ast}}M_{-}}
{(k-\xi_{n}^{\ast})^{2}}+\frac{\mathop{\mbox{Res}}\limits_{k=\xi_{n}^{\ast}}M_{-}}{(k-\xi_{n}^{\ast})}\right]-M_{+}G,\label{32a}
\end{gather}
which can be solved as follows by the Plemelj's formulae
\begin{gather}
M(x, t; k)=I+\frac{1}{2\pi i}\int_{\Sigma}\frac{M_{+}(x, t; \zeta)G(x, t; \zeta)}{\zeta-k}d\zeta\notag\\
+\sum_{n=1}^{2N}\left[\frac{\mathop{L_{-3}}\limits_{k=\xi_{n}}M_{+}}{(k-\xi_{n})^{3}}
+\frac{\mathop{L_{-2}}\limits_{k=\xi_{n}}M_{+}}{(k-\xi_{n})^{2}}
+\frac{\mathop{\mbox{Res}}\limits_{k=\xi_{n}}M_{+}}{k-\xi_{n}}+\frac{\mathop{L_{-3}}\limits_{k=\xi_{n}^{\ast}}M_{-}}
{(k-\xi_{n}^{\ast})^{3}}+\frac{\mathop{L_{-2}}\limits_{k=\xi_{n}^{\ast}}M_{-}}
{(k-\xi_{n}^{\ast})^{2}}+\frac{\mathop{\mbox{Res}}\limits_{k=\xi_{n}^{\ast}}M_{-}}{(k-\xi_{n}^{\ast})}\right].\label{33a}
\end{gather}
where
\begin{gather}
\frac{\mathop{L_{-3}}\limits_{k=\xi_{n}}M_{+}}{(k-\xi_{n})^{3}}
+\frac{\mathop{L_{-2}}\limits_{k=\xi_{n}}M_{+}}{(k-\xi_{n})^{2}}
+\frac{\mathop{\mbox{Res}}\limits_{k=\xi_{n}}M_{+}}{k-\xi_{n}}+\frac{\mathop{L_{-3}}\limits_{k=\xi_{n}^{\ast}}M_{-}}
{(k-\xi_{n}^{\ast})^{3}}+\frac{\mathop{L_{-2}}\limits_{k=\xi_{n}^{\ast}}M_{-}}
{(k-\xi_{n}^{\ast})^{2}}+\frac{\mathop{\mbox{Res}}\limits_{k=\xi_{n}^{\ast}}M_{-}}{(k-\xi_{n}^{\ast})}=\notag\\
\left(\hat{C}_{n}(k)\left[\frac{1}{2}\mu''_{-2}(\xi_{n}^{\ast})+\left(\hat{D}_{n}+\frac{1}{k-\xi_{n}^{\ast}}\right)\mu'_{-2}(\xi_{n}^{\ast})+(\frac{1}{(k-\xi^{\ast}_{n})^{2}}+\frac{\hat{D}_{n}}{k-\xi^{\ast}_{n}}+\hat{F}_{n})\mu_{-2}(\xi_{n}^{\ast})\right],\right.\notag\\
\left.C_{n}(k)\left[\frac{1}{2}\mu''_{-1}(\xi_{n})+\left(D_{n}+\frac{1}{k-\xi_{n}}\right)\mu'_{-1}(\xi_{n})+(\frac{1}{(k-\xi_{n})^{2}}+\frac{D_{n}}{k-\xi_{n}}+F_{n})\mu_{-1}(\xi_{n})\right]\right),\label{34a}
\end{gather}
and
\begin{align}\label{35a}
&C_{n}(k)=\frac{\tilde{A}[\xi_{n}]e^{-2i\theta(\xi_{n})}}{k-\xi_{n}}, \ D_{n}=\tilde{B}[\xi_{n}]-2i\theta'(\xi_{n}), \ F_{n}=\tilde{C}[\xi_{n}]-\Theta_{1}(\xi_{n}) \notag\\
&\hat{C}_{n}(k)=\frac{\tilde{A}[\xi_{n}^{\ast}]e^{2i\theta(\xi_{n}^{\ast})}}{k-\xi_{n}^{\ast}}, \ \hat{D}_{n}=\tilde{B}[\xi_{n}^{\ast}]+2i\theta'(\xi_{n}^{\ast}),\ \hat{F}_{n}=\tilde{C}[\xi^{\ast}_{n}]-\Theta_{2}(\xi^{\ast}_{n}).
\end{align}
Furthermore, according to \eqref{28}, we have
\begin{gather}
M^{(1)}(x, t; k)=-\frac{1}{2\pi i}\int_{\Sigma}M_{+}(x, t; \zeta)G(x, t; \zeta)d\zeta + \notag\\
\sum_{n=1}^{2N}\left(\tilde{A}[\xi_{n}^{\ast}]e^{2i\theta(\xi_{n}^{\ast})}\left(\frac{1}{2}\mu''_{-2}(\xi_{n}^{\ast})+\hat{D}_{n}\mu'_{-2}(\xi_{n}^{\ast})+\hat{F}_{n}\mu_{-2}(\xi_{n}^{\ast})\right),
\tilde{A}[\xi_{n}]e^{-2i\theta(\xi_{n})}\left(\frac{1}{2}\mu''_{-1}(\xi_{n})+D_{n}\mu'_{-1}(\xi_{n})+F_{n}\mu_{1}(\xi_{n})\right)\right). \label{37a}
\end{gather}
The potential $u(x, t)$ with triple poles for the GI equation with ZBCs \eqref{5} is reconstructed as
\begin{gather}
u(x, t)=2iM_{12}^{(1)}=-\frac{1}{\pi}\int_{\Sigma}(M_{+}(x, t; \zeta)G(x, t; \zeta))_{12}d\zeta \notag\\
+2i\sum_{n=1}^{2N}\tilde{A}[\xi_{n}]e^{-2i\theta(\xi_{n})}\left(\frac{1}{2}\mu''_{-11}(\xi_{n})+D_{n}\mu'_{-11}(\xi_{n})+F_{n}\mu_{11}(\xi_{n})\right)
.\label{38a}
\end{gather}

\subsection{Triple poles soliton solutions with ZBCs}
Taking $\rho(k)=\tilde{\rho}(k)=0$, we can derive the explicit triple poles soliton solutions of the GI equation with ZBCs. Then, the second column of Eq.\eqref{33a} yields
\begin{align}\label{39a}
&\mu_{-2}(\xi_{j}^{\ast})=\left(\begin{array}{c}
     0\\
  1\\
\end{array}\right)+\sum_{n=1}^{2N}C_{n}(\xi_{j}^{\ast})\left[\frac{1}{2}\mu''_{-1}(\xi_{n})+\left(D_{n}+\frac{1}{\xi_{j}^{\ast}-\xi_{n}}\right)\mu'_{-1}(\xi_{n})+(\frac{1}{(\xi_{j}^{\ast}-\xi_{n})^{2}}+\frac{D_{n}}{\xi_{j}^{\ast}-\xi_{n}}+F_{n})\mu_{-1}(\xi_{n})\right],
\notag\\
&\mu_{-2}'(\xi_{j}^{\ast})=-\sum_{n=1}^{2N}\frac{C_{n}(\xi_{j}^{\ast})}{\xi_{j}^{\ast}-\xi_{n}}\left[\frac{1}{2}\mu''_{-1}(\xi_{n})+\left(D_{n}+\frac{2}{\xi_{j}^{\ast}-\xi_{n}}\right)\mu'_{-1}(\xi_{n})+(\frac{3}{(\xi_{j}^{\ast}-\xi_{n})^{2}}+\frac{2D_{n}}{\xi_{j}^{\ast}-\xi_{n}}+F_{n})\mu_{-1}(\xi_{n})\right],
\notag\\
&\mu_{-2}''(\xi_{j}^{\ast})=\sum_{n=1}^{2N}\frac{2C_{n}(\xi_{j}^{\ast})}{(\xi_{j}^{\ast}-\xi_{n})^{2}}\left[\frac{1}{2}\mu''_{-1}(\xi_{n})+\left(D_{n}+\frac{3}{\xi_{j}^{\ast}-\xi_{n}}\right)\mu'_{-1}(\xi_{n})+(\frac{6}{(\xi_{j}^{\ast}-\xi_{n})^{2}}+\frac{3D_{n}}{\xi_{j}^{\ast}-\xi_{n}}+F_{n})\mu_{-1}(\xi_{n})\right],
\notag\\
&\mu_{-1}(\xi_{j})=\left(\begin{array}{c}
     1\\
  0\\
\end{array}\right)+\sum_{n=1}^{2N}\hat{C}_{n}(\xi_{j})\left[\frac{1}{2}\mu''_{-2}(\xi_{n}^{\ast})+\left(\hat{D}_{n}+\frac{1}{\xi_{j}-\xi_{n}^{\ast}}\right)\mu'_{-2}(\xi_{n}^{\ast})+(\frac{1}{(\xi_{j}-\xi^{\ast}_{n})^{2}}+\frac{\hat{D}_{n}}{\xi_{j}-\xi^{\ast}_{n}}+\hat{F}_{n})\mu_{-2}(\xi_{n}^{\ast})\right],
\notag\\
&\mu_{-1}'(\xi_{j})=-\sum_{n=1}^{2N}\frac{\hat{C}_{n}(\xi_{j})}{\xi_{j}-\xi_{n}^{\ast}}\left[\frac{1}{2}\mu''_{-2}(\xi_{n}^{\ast})+\left(\hat{D}_{n}+\frac{2}{\xi_{j}-\xi_{n}^{\ast}}\right)\mu'_{-2}(\xi_{n}^{\ast})+(\frac{3}{(\xi_{j}-\xi^{\ast}_{n})^{2}}+\frac{2\hat{D}_{n}}{\xi_{j}-\xi^{\ast}_{n}}+\hat{F}_{n})\mu_{-2}(\xi_{n}^{\ast})\right],\notag\\
&\mu_{-1}''(\xi_{j})=\sum_{n=1}^{2N}\frac{2\hat{C}_{n}(\xi_{j})}{(\xi_{j}-\xi_{n}^{\ast})^{2}}\left[\frac{1}{2}\mu''_{-2}(\xi_{n}^{\ast})+\left(\hat{D}_{n}+\frac{3}{\xi_{j}-\xi_{n}^{\ast}}\right)\mu'_{-2}(\xi_{n}^{\ast})+(\frac{6}{(\xi_{j}-\xi^{\ast}_{n})^{2}}+\frac{3\hat{D}_{n}}{\xi_{j}-\xi^{\ast}_{n}}+\hat{F}_{n})\mu_{-2}(\xi_{n}^{\ast})\right].
\end{align}
Furthermore, we have
\begin{align}\label{40a}
&\mu_{-12}(\xi_{j}^{\ast})=\sum_{n=1}^{2N}C_{n}(\xi_{j}^{\ast})\left[\frac{1}{2}\mu''_{-11}(\xi_{n})+\left(D_{n}+\frac{1}{\xi_{j}^{\ast}-\xi_{n}}\right)\mu'_{-11}(\xi_{n})+(\frac{1}{(\xi_{j}^{\ast}-\xi_{n})^{2}}+\frac{D_{n}}{\xi_{j}^{\ast}-\xi_{n}}+F_{n})\mu_{-11}(\xi_{n})\right],
\notag\\
&\mu_{-12}'(\xi_{j}^{\ast})=-\sum_{n=1}^{2N}\frac{C_{n}(\xi_{j}^{\ast})}{\xi_{j}^{\ast}-\xi_{n}}\left[\frac{1}{2}\mu''_{-11}(\xi_{n})+\left(D_{n}+\frac{2}{\xi_{j}^{\ast}-\xi_{n}}\right)\mu'_{-11}(\xi_{n})+(\frac{3}{(\xi_{j}^{\ast}-\xi_{n})^{2}}+\frac{2D_{n}}{\xi_{j}^{\ast}-\xi_{n}}+F_{n})\mu_{-11}(\xi_{n})\right],
\notag\\
&\mu_{-12}''(\xi_{j}^{\ast})=\sum_{n=1}^{2N}\frac{2C_{n}(\xi_{j}^{\ast})}{(\xi_{j}^{\ast}-\xi_{n})^{2}}\left[\frac{1}{2}\mu''_{-11}(\xi_{n})+\left(D_{n}+\frac{3}{\xi_{j}^{\ast}-\xi_{n}}\right)\mu'_{-11}(\xi_{n})+(\frac{6}{(\xi_{j}^{\ast}-\xi_{n})^{2}}+\frac{3D_{n}}{\xi_{j}^{\ast}-\xi_{n}}+F_{n})\mu_{-11}(\xi_{n})\right],
\notag\\
&\mu_{-11}(\xi_{j})=1+\sum_{n=1}^{2N}\hat{C}_{n}(\xi_{j})\left[\frac{1}{2}\mu''_{-12}(\xi_{n}^{\ast})+\left(\hat{D}_{n}+\frac{1}{\xi_{j}-\xi_{n}^{\ast}}\right)\mu'_{-12}(\xi_{n}^{\ast})+(\frac{1}{(\xi_{j}-\xi^{\ast}_{n})^{2}}+\frac{\hat{D}_{n}}{\xi_{j}-\xi^{\ast}_{n}}+\hat{F}_{n})\mu_{-12}(\xi_{n}^{\ast})\right],
\notag\\
&\mu_{-11}'(\xi_{j})=-\sum_{n=1}^{2N}\frac{\hat{C}_{n}(\xi_{j})}{\xi_{j}-\xi_{n}^{\ast}}\left[\frac{1}{2}\mu''_{-12}(\xi_{n}^{\ast})+\left(\hat{D}_{n}+\frac{2}{\xi_{j}-\xi_{n}^{\ast}}\right)\mu'_{-12}(\xi_{n}^{\ast})+(\frac{3}{(\xi_{j}-\xi^{\ast}_{n})^{2}}+\frac{2\hat{D}_{n}}{\xi_{j}-\xi^{\ast}_{n}}+\hat{F}_{n})\mu_{-12}(\xi_{n}^{\ast})\right],\notag\\
&\mu_{-11}''(\xi_{j})=\sum_{n=1}^{2N}\frac{2\hat{C}_{n}(\xi_{j})}{(\xi_{j}-\xi_{n}^{\ast})^{2}}\left[\frac{1}{2}\mu''_{-12}(\xi_{n}^{\ast})+\left(\hat{D}_{n}+\frac{3}{\xi_{j}-\xi_{n}^{\ast}}\right)\mu'_{-12}(\xi_{n}^{\ast})+(\frac{6}{(\xi_{j}-\xi^{\ast}_{n})^{2}}+\frac{3\hat{D}_{n}}{\xi_{j}-\xi^{\ast}_{n}}+\hat{F}_{n})\mu_{-12}(\xi_{n}^{\ast})\right].
\end{align}
\noindent \textbf{Theorem 4.1}  \emph{
The explicit formula of the  triple poles soliton solution for the GI equation \eqref{1} with ZBCs \eqref{5} is expressed as
\begin{align}\label{41a}
u(x,t)=-2i\frac{\det \left(\begin{array}{cc}
    I-\tilde{H}  &  \tilde{\beta}\\
  \tilde{\alpha}^{T} &  0\\
\end{array}\right)}{\det (I-\tilde{H})},
\end{align}
where $\tilde{\beta}, \tilde{H}, \tilde{\alpha}$ are given in \eqref{43.1a}, \eqref{42a} and \eqref{44.1a}.
}
\begin{proof}
We rewrite the  linear system  \eqref{40a}  in the matrix form:
\begin{align}\label{43a}
\tilde{\psi}-\tilde{H}\tilde{\psi}=\tilde{\beta},
\end{align}
where
\begin{gather}
\tilde{\psi}=\left(\begin{array}{c}
     \tilde{\psi}^{(1)}\\
  \tilde{\psi}^{(2)}\\
  \tilde{\psi}^{(3)}\\
\end{array}\right), \tilde{\psi}^{(1)}=(\mu''_{-11}(\xi_{1}),\cdots, \mu''_{-11}(\xi_{2N}))^{T},  \tilde{\psi}^{(2)}=(\mu'_{-11}(\xi_{1}),\cdots, \mu'_{-11}(\xi_{2N}))^{T},\notag\\ \tilde{\psi}^{(3)}=(\mu_{-11}(\xi_{1}),\cdots, \mu_{-11}(\xi_{2N}))^{T},
 \tilde{\beta}=\left(\begin{array}{c}
     \tilde{\beta}^{(1)}\\
  \tilde{\beta}^{(2)}\\
  \tilde{\beta}^{(3)}\\
\end{array}\right), \tilde{\beta}^{(1)}=(0)_{2N\times 1}, \tilde{\beta}^{(2)}=(0)_{2N\times 1}, \tilde{\beta}^{(3)}=(1)_{2N\times 1}. \label{43.1a}
\end{gather}
The $6N\times 6N$ matrix $\tilde{H}=\left(\tilde{H}^{(im)}\right)_{3\times 3}$ with $\tilde{H}^{(im)}=\left(\tilde{H}^{(im)}_{jk}\right)_{2N\times 2N}(i, m=
1, 2, 3)$ given by
\begin{gather}
\tilde{H}^{(11)}_{jk}=\sum_{n=1}^{2N}\frac{\hat{C}_{n}(\xi_{j})C_{k}(\xi_{n}^{\ast})}{(\xi_{j}-\xi_{n}^{\ast})^{2}}\left[\frac{1}{(\xi_{n}^{\ast}-\xi_{k})^{2}}-\frac{1}{\xi_{n}^{\ast}-\xi_{k}}
(\hat{D}_{n}+\frac{3}{\xi_{j}-\xi_{n}^{\ast}})+(\frac{6}{(\xi_{j}-\xi^{\ast}_{n})^{2}}+\frac{3\hat{D}_{n}}{\xi_{j}-\xi^{\ast}_{n}}+\hat{F}_{n})\right],\notag\\
\tilde{H}^{(12)}_{jk}=\sum_{n=1}^{2N}\frac{2\hat{C}_{n}(\xi_{j})C_{k}(\xi_{n}^{\ast})}{(\xi_{j}-\xi_{n}^{\ast})^{2}}\left[\frac{1}{(\xi_{n}^{\ast}-\xi_{k})^{2}}(D_{k}+\frac{3}{\xi_{n}^{\ast}-\xi_{k}})-\frac{1}{\xi_{n}^{\ast}-\xi_{k}}
(\hat{D}_{n}+\frac{3}{\xi_{j}-\xi_{n}^{\ast}})(D_{k}+\frac{2}{\xi_{n}^{\ast}-\xi_{k}})\right.\notag\\
\left.+(\frac{6}{(\xi_{j}-\xi^{\ast}_{n})^{2}}+\frac{3\hat{D}_{n}}{\xi_{j}-\xi^{\ast}_{n}}+\hat{F}_{n})(D_{k}+\frac{1}{\xi_{n}^{\ast}-\xi_{k}})\right],\notag\\
\tilde{H}^{(13)}_{jk}=\sum_{n=1}^{2N}\frac{2\hat{C}_{n}(\xi_{j})C_{k}(\xi_{n}^{\ast})}{(\xi_{j}-\xi_{n}^{\ast})^{2}}\left[\frac{1}{(\xi_{n}^{\ast}-\xi_{k})^{2}}(\frac{6}{(\xi^{\ast}_{n}-\xi_{k})^{2}}+\frac{3D_{k}}{\xi^{\ast}_{n}-\xi_{k}}+F_{k})\right.\notag\\
\left.-\frac{1}{\xi_{n}^{\ast}-\xi_{k}}
(\hat{D}_{n}+\frac{3}{\xi_{j}-\xi_{n}^{\ast}})(\frac{3}{(\xi^{\ast}_{n}-\xi_{k})^{2}}+\frac{2D_{k}}{\xi^{\ast}_{n}-\xi_{k}}+F_{k})\right.\notag\\
\left.+(\frac{6}{(\xi_{j}-\xi^{\ast}_{n})^{2}}+\frac{3\hat{D}_{n}}{\xi_{j}-\xi^{\ast}_{n}}+\hat{F}_{n})(\frac{1}{(\xi^{\ast}_{n}-\xi_{k})^{2}}+\frac{D_{k}}{\xi^{\ast}_{n}-\xi_{k}}+F_{k})\right],\label{42a}
\end{gather}
\begin{gather}
\tilde{H}^{(21)}_{jk}=-\sum_{n=1}^{2N}\frac{\hat{C}_{n}(\xi_{j})C_{k}(\xi_{n}^{\ast})}{2(\xi_{j}-\xi_{n}^{\ast})}\left[\frac{1}{(\xi_{n}^{\ast}-\xi_{k})^{2}}-\frac{1}{\xi_{n}^{\ast}-\xi_{k}}
(\hat{D}_{n}+\frac{2}{\xi_{j}-\xi_{n}^{\ast}})+(\frac{3}{(\xi_{j}-\xi^{\ast}_{n})^{2}}+\frac{2\hat{D}_{n}}{\xi_{j}-\xi^{\ast}_{n}}+\hat{F}_{n})\right],\notag\\
\tilde{H}^{(22)}_{jk}=-\sum_{n=1}^{2N}\frac{\hat{C}_{n}(\xi_{j})C_{k}(\xi_{n}^{\ast})}{\xi_{j}-\xi_{n}^{\ast}}\left[\frac{1}{(\xi_{n}^{\ast}-\xi_{k})^{2}}(D_{k}+\frac{3}{\xi_{n}^{\ast}-\xi_{k}})-\frac{1}{\xi_{n}^{\ast}-\xi_{k}}
(\hat{D}_{n}+\frac{2}{\xi_{j}-\xi_{n}^{\ast}})(D_{k}+\frac{2}{\xi_{n}^{\ast}-\xi_{k}})\right.\notag\\
\left.+(\frac{3}{(\xi_{j}-\xi^{\ast}_{n})^{2}}+\frac{2\hat{D}_{n}}{\xi_{j}-\xi^{\ast}_{n}}+\hat{F}_{n})(D_{k}+\frac{1}{\xi_{n}^{\ast}-\xi_{k}})\right],\notag\\
\tilde{H}^{(23)}_{jk}=-\sum_{n=1}^{2N}\frac{\hat{C}_{n}(\xi_{j})C_{k}(\xi_{n}^{\ast})}{\xi_{j}-\xi_{n}^{\ast}}\left[\frac{1}{(\xi_{n}^{\ast}-\xi_{k})^{2}}(\frac{6}{(\xi^{\ast}_{n}-\xi_{k})^{2}}+\frac{3D_{k}
}{\xi^{\ast}_{n}-\xi_{k}}+F_{k})\right.\notag\\
\left.-\frac{1}{\xi_{n}^{\ast}-\xi_{k}}
(\hat{D}_{n}+\frac{2}{\xi_{j}-\xi_{n}^{\ast}})(\frac{3}{(\xi^{\ast}_{n}-\xi_{k})^{2}}+\frac{2D_{k}}{\xi^{\ast}_{n}-\xi_{k}}+F_{k})\right.\notag\\
\left.+(\frac{3}{(\xi_{j}-\xi^{\ast}_{n})^{2}}+\frac{2\hat{D}_{n}}{\xi_{j}-\xi^{\ast}_{n}}+\hat{F}_{n})(\frac{1}{(\xi^{\ast}_{n}-\xi_{k})^{2}}+\frac{D_{k}
}{\xi^{\ast}_{n}-\xi_{k}}+F_{k})\right],\label{42a1}
\end{gather}
\begin{gather}
\tilde{H}^{(31)}_{jk}=\sum_{n=1}^{2N}\frac{\hat{C}_{n}(\xi_{j})C_{k}(\xi_{n}^{\ast})}{2}\left[\frac{1}{(\xi_{n}^{\ast}-\xi_{k})^{2}}-\frac{1}{\xi_{n}^{\ast}-\xi_{k}}
(\hat{D}_{n}+\frac{1}{\xi_{j}-\xi_{n}^{\ast}})+(\frac{1}{(\xi_{j}-\xi^{\ast}_{n})^{2}}+\frac{\hat{D}_{n}}{\xi_{j}-\xi^{\ast}_{n}}+\hat{F}_{n})\right],\notag\\
\tilde{H}^{(32)}_{jk}=\sum_{n=1}^{2N}\hat{C}_{n}(\xi_{j})C_{k}(\xi_{n}^{\ast})\left[\frac{1}{(\xi_{n}^{\ast}-\xi_{k})^{2}}(D_{k}+\frac{3}{\xi_{n}^{\ast}-\xi_{k}})-\frac{1}{\xi_{n}^{\ast}-\xi_{k}}
(\hat{D}_{n}+\frac{1}{\xi_{j}-\xi_{n}^{\ast}})(D_{k}+\frac{2}{\xi_{n}^{\ast}-\xi_{k}})\right.\notag\\
\left.+(\frac{1}{(\xi_{j}-\xi^{\ast}_{n})^{2}}+\frac{\hat{D}_{n}}{\xi_{j}-\xi^{\ast}_{n}}+\hat{F}_{n})(D_{k}+\frac{1}{\xi_{n}^{\ast}-\xi_{k}})\right],\notag\\
\tilde{H}^{(33)}_{jk}=\sum_{n=1}^{2N}\hat{C}_{n}(\xi_{j})C_{k}(\xi_{n}^{\ast})\left[\frac{1}{(\xi_{n}^{\ast}-\xi_{k})^{2}}(\frac{6}{(\xi^{\ast}_{n}-\xi_{k})^{2}}+\frac{3D_{k}
}{\xi^{\ast}_{n}-\xi_{k}}+F_{k})\right.\notag\\
\left.-\frac{1}{\xi_{n}^{\ast}-\xi_{k}}
(\hat{D}_{n}+\frac{1}{\xi_{j}-\xi_{n}^{\ast}})(\frac{3}{(\xi^{\ast}_{n}-\xi_{k})^{2}}+\frac{2D_{k}}{\xi^{\ast}_{n}-\xi_{k}}+F_{k})\right.\notag\\
\left.+(\frac{1}{(\xi_{j}-\xi^{\ast}_{n})^{2}}+\frac{\hat{D}_{n}}{\xi_{j}-\xi^{\ast}_{n}}+\hat{F}_{n})(\frac{1}{(\xi^{\ast}_{n}-\xi_{k})^{2}}+\frac{D_{k}
}{\xi^{\ast}_{n}-\xi_{k}}+F_{k})\right],\label{42a2}
\end{gather}
At the case of reflectionless potential, Eq.\eqref{38a} can be redefined as
\begin{align}\label{44a}
u=2i\tilde{\alpha}^{T} \tilde{\psi},
\end{align}
where
\begin{gather}
\tilde{\alpha}=\left(\begin{array}{c}
     \tilde{\alpha}^{(1)}\\
  \tilde{\alpha}^{(2)}\\
   \tilde{\alpha}^{(3)}\\
\end{array}\right),   \tilde{\alpha}^{(1)}=(\frac{1}{2}\tilde{A}[\xi_{1}]e^{-2i\theta(\xi_{1})}, \frac{1}{2}\tilde{A}[\xi_{2}]e^{-2i\theta(\xi_{2})},\cdots, \frac{1}{2}\tilde{A}[\xi_{2N}]e^{-2i\theta(\xi_{2N})})^{T},\notag\\
\tilde{\alpha}^{(2)}=(\tilde{A}[\xi_{1}]e^{-2i\theta(\xi_{1})}D_{1}, \tilde{A}[\xi_{2}]e^{-2i\theta(\xi_{2})}D_{2},\cdots, \tilde{A}[\xi_{2N}]e^{-2i\theta(\xi_{2N})}D_{2N})^{T},\notag\\
\tilde{\alpha}^{(3)}=(\tilde{A}[\xi_{1}]e^{-2i\theta(\xi_{1})}F_{1}, \tilde{A}[\xi_{2}]e^{-2i\theta(\xi_{2})}F_{2},\cdots, \tilde{A}[\xi_{2N}]e^{-2i\theta(\xi_{2N})}F_{2N})^{T}.\label{44.1a}
\end{gather}
Combining Eqs. \eqref{43a}, the triple poles soliton solution \eqref{41a} can be given out.
\end{proof}

As a matter of convenience, we take $N=1$ as a example to illustrate the correlative dynamic behavior for the one-triple poles soliton solution for GI equation with ZBCs \eqref{5}. As we can see in Fig. 4, it displays the bright-bright-bright soliton solutions, which stands for the interaction of three bright soliton waves.\\
{\rotatebox{0}{\includegraphics[width=5.0cm,height=5.0cm,angle=0]{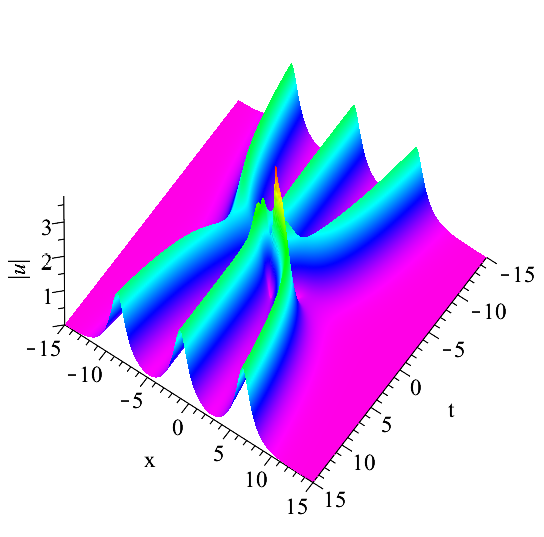}}}
~~~~\quad\qquad
{\rotatebox{0}{\includegraphics[width=5.0cm,height=5.0cm,angle=0]{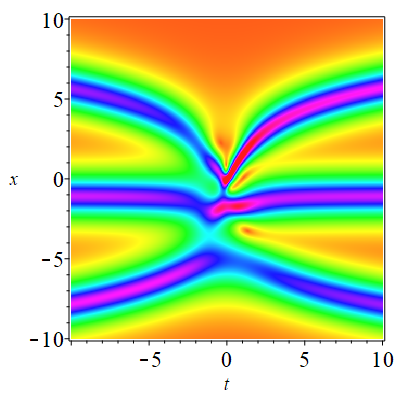}}}
~~~~\quad\qquad
{\rotatebox{0}{\includegraphics[width=5.0cm,height=5.0cm,angle=0]{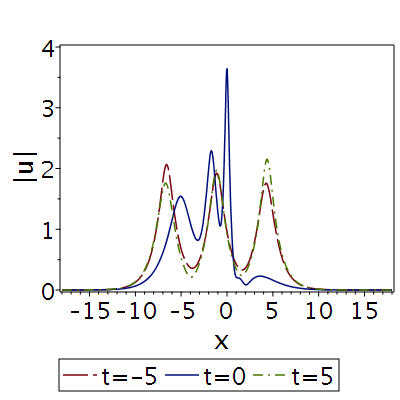}}}

$\qquad\qquad\textbf{(a)}\qquad\qquad\qquad\qquad\qquad\qquad\quad\qquad\qquad\quad 
\textbf{(b)}\qquad\qquad\qquad\qquad\qquad\qquad\qquad\qquad\quad\textbf{(c)}$\\
\noindent { \small \textbf{Figure 4.} (Color online) The one-triple pole soliton solution for Eq.\eqref{1} with ZBCs and $N=1$. The parameters are $A[k_{1}]=B[k_{1}]=C[k_{1}]=1, k_{1}=\frac{1}{2}+\frac{1}{2}i$. $\textbf{(a)}$ Three dimensional plot;
$\textbf{(b)}$ The density plot;
$\textbf{(c)}$ The wave propagation along the $x$-axis at $t=-5$(longdash), $t=0$(solid), $t=5$(dashdot).}\\

\section{The construction of Riemann-Hilbert problem with NZBCs}
The direct scattering problem for the GI equation \eqref{1} with
NZBCs  has been studied in Ref.\cite{Fan-GI39,Fan-GI40}. In this section, we first recall some results for the direct scattering
problem for the targeted GI equation with following NZBCs at infinity
\begin{align}\label{49}
\lim_{x\rightarrow \pm \infty}u(x, t)=u_{\pm}e^{-\frac{3}{2}iu_{0}^{4}t+iu_{0}^{2}x},
\end{align}
where $|u_{\pm}|=u_{0}>0$, and $u_{\pm}$ are constant.

\subsection{Spectral analysis}
As $x \rightarrow \pm \infty$, the Lax pair \eqref{2} under the boundary \eqref{49} becomes
\begin{align}\label{50}
\Phi_{x}=X_{\pm}\Phi=-ik^{2}\sigma_{3}+kQ_{\pm}, \qquad \Phi_{t}=T_{\pm}\Phi=(2k^{2}-u_{0}^{2})X_{\pm},
\end{align}
where
\begin{align}\label{51}
Q_{\pm}=\left(\begin{array}{cc}
    0  &  u_{\pm}\\
    -u_{\pm}^{\ast} &  0\\
\end{array}\right).
\end{align}
The fundamental matrix solution of this lax pair is
\begin{align}\label{52}
\Phi_{\pm}^{bg}(x, t; k)=\left\{
\begin{array}{lr}
Y_{\pm}(k)e^{-i\theta(x, t; k)\sigma_{3}}, \quad k\neq\pm iu_{0},\\
I+(x-3u_{0}^{2}t)X_{\pm}(k), \quad k=\pm iu_{0},
\end{array}
\right.
\end{align}
where
\begin{align}\label{53}
Y_{\pm}=\left(\begin{array}{cc}
    1  &  -\frac{iu_{\pm}}{\lambda+k}\\
    -\frac{iu_{\pm}^{\ast}}{\lambda+k} &  1\\
\end{array}\right),\quad \theta(x, t; k)=k\lambda[x+(2k^{2}-u_{0}^{2})t], \quad \lambda^{2}=k^{2}+u_{0}^{2}.
\end{align}
Introducing a uniformization variable $z=k+\lambda$, we obtain
\begin{align}\label{54}
k=\frac{1}{2}(z-\frac{u_{0}^{2}}{z}), \qquad \lambda=\frac{1}{2}(z+\frac{u_{0}^{2}}{z}),
\end{align}
which means that the scattering problem can be analysed on a standard $z$-plane instead of the two-sheeted
Riemann surface.

Defining $D_{+}$, $D_{-}$ and $\Sigma$ on $z$-plane as $\Sigma=\mathbb{R}\cup i\mathbb{R}\setminus \{0\}, D_{\pm}= \left\{z\in \mathbb{C} | \mbox{Re}z\mbox{Im}z\gtrless 0\right\},$ the Jost solutions $\Phi_{\pm}(x, t,; z)$ are given by
\begin{align}\label{56}
\Phi_{\pm}(x, t; z)=Y_{\pm}e^{-i\theta(x, t; z)\sigma_{3}}+o(1), \quad z\in \Sigma, \quad \mbox{as} \quad x\rightarrow \pm \infty.
\end{align}
After the variable transformation
\begin{align}\label{57}
\mu_{\pm}(x, t; z)=\Phi_{\pm}(x, t; z)e^{i\theta(x, t; z)\sigma_{3}},
\end{align}
the modified Jost solutions $\mu_{\pm}(x, t; z)$ tend to $Y_{\pm}(z)$ as $x\rightarrow \pm\infty$, and it also satisfy the following Volterra integral equations
\begin{align}\label{58}
\mu_{\pm}(x,t,z)=\left\{
\begin{array}{lr}
Y_{\pm}+\int_{\pm\infty}^{x}Y_{\pm}e^{-ik\lambda(x-y)\hat{\sigma}_{3}}
\left[Y_{\pm}^{-1}\Delta X_{\pm}(y, t)\mu_{\pm}(y,t;z)\right]dy, \quad z\neq\pm iu_{0},\\
Y_{\pm}+\int_{\pm\infty}^{x}
\left[I+(x-y)X_{\pm}(z)\right]\Delta X_{\pm}(y, t)\mu_{\pm}(y,t;z)dy, \quad z=\pm iu_{0},
\end{array}
\right.
\end{align}
where $\Delta X_{\pm}=X-X_{\pm}$.

\noindent \textbf{Proposition 5.1.} \emph{
Suppose $u-u_{\pm}\in L^{1} (\mathbb{R}^{\pm})$, then $\mu_{\pm}(x,t;z)$ given in Eq.\eqref{57} are unique solutions for the Jost integral equation \eqref{58} in $\Sigma_{0}=\Sigma\setminus\{\pm iu_{0}\}$, $\mu_{\pm}(x,t;z)$ have the following characteristics:}\\
\emph{$\bullet$ $\mu_{-1}(x, t; z)$ and $\mu_{+2}(x, t; z)$ become analytical for $D_{+}$ and
continuous in $D_{+}\cup \Sigma_{0}$;}\\
\emph{$\bullet$ $\mu_{+1}(x, t; z)$ and $\mu_{-2}(x, t; z)$  become analytical for $D_{-}$ and
continuous in $D_{-}\cup \Sigma_{0}$;}\\
\emph{$\bullet$ $\mu_{\pm}(x,t;z)\rightarrow I$ \mbox{as}  $z\rightarrow \infty$;}\\
\emph{$\bullet$ $\mu_{\pm}(x,t;z) \rightarrow -\frac{i}{z}\sigma_{3}Q_{\pm}$ \mbox{as}  $z\rightarrow 0$;}\\
\emph{$\bullet$ $\det \mu_{\pm}(x, t; z)=\det Y_{\pm}=\gamma=1+\frac{u_{0}^{2}}{z^{2}}, \quad x, t\in \mathbb{R}, \quad z\in \Sigma_{0}$.}

\centerline{\begin{tikzpicture}[scale=1.8]
\path [fill=gray] (2.5,0) -- (0.5,0) to
(0.5,-2) -- (2.5,-2);
\path [fill=gray] (4.5,0) -- (2.5,0) to
(2.5,2) -- (4.5,2);
\draw[-][thick](0.5,0)--(0.75,0);
\draw[<-][thick](0.75,0)--(1,0);
\draw[-][thick](1,0)--(2,0);
\draw[<-][thick](2,0)--(2.5,0);
\draw[fill] (2.5,0) circle [radius=0.03];
\draw[->][thick](2.5,0)--(3,0);
\draw[->][thick](3,0)--(4,0);
\draw[-][thick](4,0)--(4.5,0)node[above]{$\mbox{Re}z$};
\draw[-][thick](2.5,2)node[right]{$\mbox{Im}z$}--(2.5,0);
\draw[-][thick](2.5,0)--(2.5,-2);
\draw[->][thick](2.5,-1.5)--(2.5,-0.5);
\draw[->][thick](2.5,-2)--(2.5,-1.5);
\draw[->][thick](2.5,1.5)--(2.5,0.5);
\draw[->][thick](2.5,2)--(2.5,1.5);
\draw[fill] (2.5,-0.3) node[right]{$0$};
\draw[fill] (2.5,1) circle [radius=0.03];
\draw[fill] (2.5,-1) circle [radius=0.03];
\draw[fill](3.8,1.5) circle [radius=0.03] node[right]{$z_{n}$};
\draw[fill] (3.8,-1.5) circle [radius=0.03] node[right]{$z^{*}_{n}$};
\draw[fill] (1.2,1.5) circle [radius=0.03] node[left]{$-z^{*}_{n}$};
\draw[fill] (1.2,-1.5) circle [radius=0.03] node[left]{$-z_{n}$};
\draw[fill] (2,0.5) circle [radius=0.03] node[right]{$-\frac{u^{2}_{0}}{z_{n}}$};
\draw[fill] (2,-0.5) circle [radius=0.03] node[right]{$-\frac{u^{2}_{0}}{z^{*}_{n}}$};
\draw[fill] (3,0.5) circle [radius=0.03] node[left]{$\frac{u^{2}_{0}}{z^{*}_{n}}$};
\draw[fill] (3,-0.5) circle [radius=0.03] node[left]{$\frac{u^{2}_{0}}{z_{n}}$};
\draw[-][thick](3.5,0) arc(0:360:1);
\draw[-][thick](3.5,0) arc(0:30:1);
\draw[-][thick](3.5,0) arc(0:150:1);
\draw[-][thick](3.5,0) arc(0:210:1);
\draw[-][thick](3.5,0) arc(0:330:1);
\draw[fill] (2,0.854) circle [radius=0.03];
\draw[fill] (1.9,0.852) circle [radius=0.0] node[above]{$-\omega^{*}_{m}$};
\draw[fill] (2,-0.854) circle [radius=0.03];
\draw[fill] (1.9,-0.851) circle [radius=0.0] node[below]{$-\omega_{m}$};
\draw[fill] (3,0.854) circle [radius=0.03];
\draw[fill] (3.1,0.851) circle [radius=0.0] node[above]{$\omega_{m}$};
\draw[fill] (3,-0.854) circle [radius=0.03];
\draw[fill] (3.1,-0.851) circle [radius=0.0] node[below]{$\omega^{*}_{m}$};
\end{tikzpicture}}
\noindent { \small \textbf{Figure 5.} (Color online) Distribution of the discrete spectrum and jumping curves for the RHP on complex $z$-plane, Region $D_{+}=\left\{z\in \mathbb{C} | \mbox{Re}z\mbox{Im}z> 0\right\}$ (gray region), region $D_{-}=\left\{z\in \mathbb{C} | \mbox{Re}z\mbox{Im}z< 0\right\}$ (white region).}\\

Due to the Jost solutions $\Phi_{\pm}(x, t; z)$ are the simultaneous solutions of spectral problem \eqref{2}, which satisfies following linear relation by the constant scattering matrix $S(z)=(s_{i j} (z))_{2\times 2}$
\begin{align}\label{59}
\Phi_{+}(x, t; z)=\Phi_{-}(x, t; z)S(z), \quad z\in \Sigma_{0},
\end{align}
where $S(z)=\sigma_{2}S^{\ast}(z^{\ast})\sigma_{2}, S(z)=\sigma_{1}S^{\ast}(-z^{\ast})\sigma_{1}, S(z)=(\sigma_{3}Q_{-})^{-1}S(-\frac{u_{0}^{2}}{z})\sigma_{3}Q_{+}$ for $z\in \Sigma$.
The scattering coefficients can be expressed as Wronskians determinant in the following form:
\begin{align}\label{60}
s_{11}(z)=\frac{Wr(\Phi_{+,1},\Phi_{-,2})}{\gamma(z)}, \quad s_{12}(z)=\frac{Wr(\Phi_{+,2},\Phi_{-,2})}{\gamma(z)},\notag\\
s_{21}(z)=\frac{Wr(\Phi_{-,1},\Phi_{+,1})}{\gamma(z)}, \quad s_{22}(z)=\frac{Wr(\Phi_{-,1},\Phi_{+,2})}{\gamma(z)}.
\end{align}

\noindent \textbf{Proposition 5.2.} \emph{ The scattering  matrix $S(z)$ satisfies:
}\\
\emph{$\bullet$ $\det S(z)=1$ for $z\in \Sigma_{0}$;}\\
\emph{$\bullet$ $s_{22}(z)$ becomes analytical  $D_{+}$ and
continuous in $D_{+}\cup \Sigma_{0}$;}\\
\emph{$\bullet$ $s_{11}(z)$ becomes analytical  $D_{-}$ and
continuous in $D_{-}\cup \Sigma_{0}$;}\\
\emph{$\bullet$ $S(x,t;z)\rightarrow I$ \mbox{as}  $z\rightarrow \infty$;}\\
\emph{$\bullet$ $S(x,t;z) \rightarrow \mbox{diag}\left(\frac{u_{-}}{u_{+}}, \frac{u_{+}}{u_{-}}\right)$ \mbox{as}  $z\rightarrow 0$.}

\subsection{The Riemann-Hilbert problem}
Based on the analytic properties of Jost solutions $\mu_{\pm}(x; t; z)$ in Proposition 5.1, we
get the following sectionally meromorphic matrices
\begin{align}\label{73}
M_{-}(x, t; z)=(\frac{\mu_{+,1}}{s_{11}},\mu_{-,2}),\qquad M_{+}(x, t; z)=(\mu_{-,1},\frac{\mu_{+,2}}{s_{22}}),
\end{align}
where superscripts $\pm$ imply analyticity in $D_{+}$ and $D_{-}$, respectively. Then, a matrix RHP is raised:

\noindent \textbf{Riemann-Hilbert Problem 2}  \emph{
$M(x, t; z)$ solves the following RHP:
\begin{align}\label{74}
\left\{
\begin{array}{lr}
M(x, t; z)\ \mbox{is analytic in} \ \mathbb{C }\setminus \Sigma,\\
M_{-}(x, t; z)=M_{+}(x, t; z)(I-G(x, t; z)), \qquad z\in \Sigma,\\
M(x, t; z)\rightarrow I,\qquad z\rightarrow \infty,\\
M(x, t; z)\rightarrow -\frac{i}{z}\sigma_{3}Q_{-},\qquad z\rightarrow 0,
  \end{array}
\right.
\end{align}
of which the jump matrix $G(x, t; z)$ is
\begin{align}\label{75}
G=\left(\begin{array}{cc}
    \rho(z)\tilde{\rho}(z)  &  e^{-2i\theta(x, t; z)}\tilde{\rho}(z)\\
  -e^{2i\theta(x, t; z)}\rho(z) &  0\\
\end{array}\right),
\end{align}
where $\rho(z)=\frac{s_{21}(z)}{s_{11}(z)}, \tilde{\rho}(z)=\frac{s_{12}(z)}{s_{22}(z)}$.
}
Taking
\begin{align}\label{76}
M(x, t; z)=I+\frac{1}{z}M^{(1)}(x, t; z)+O(\frac{1}{z^{2}}),\qquad z\rightarrow \infty,
\end{align}
then the potential $u(x, t)$ of the GI equation \eqref{1} with NZBCs is given by
\begin{align}\label{76.1}
u(x, t)=iM_{12}^{(1)}(x, t; z)=i\lim_{z\rightarrow\infty}zM_{12}(x, t; z).
\end{align}

\section{The solution of GI equation under NZBCs with double poles}
In this section, the inverse scattering problem with double poles discrete spectrum for the GI equation \eqref{1} under NZBCs will be considered, and the general $N$-double poles solutions will be given.
\subsection{Inverse scattering problem with NZBCs and double poles}
We first suppose that $s_{22}(z)$ has $N_{1}$ double zeros $z_{n}$ ($n=1, 2, \cdots, N_{1}$) in $D_{+}\cap\left\{z\in\mathbb{ C}: \mbox{Re} z>0,\right.$ $ \left.\mbox{Im} z>0, |z|>u_{0}\right\}$,  and $N_{2}$ double zeros $\omega_{m}$ in $\{z=u_{0}e^{i\vartheta}: 0<\vartheta<\frac{\pi}{2}\}$, which denotes $s_{22}(z_{0})=s'_{22}(z_{0})=0$ and $s''_{22}(z_{0})\neq 0$ when $z_{0}$ is the double zero of $s_{22}(z)$. From the symmetries of the scattering matrix, the corresponding discrete spectrum is summed up as(see Fig. 5)
\begin{align}\label{67}
\Upsilon=\left\{\pm z_{n}, \pm z_{n}^{\ast}, \pm\frac{u_{0}^{2}}{z_{n}}, \pm\frac{u_{0}^{2}}{z_{n}^{\ast}}\right\}_{n=1}^{N_{1}}\cup\{\pm\omega_{m}, \pm\omega^{\ast}_{m}\}_{m=1}^{N_{2}}.
\end{align}

Due to $s_{22}(z_{0})=0$ ($z_{0}\in \Upsilon\cap D_{+}$), we can find that $\Phi_{-1}(x, t; z_{0})$ and $\Phi_{+2}(x, t; z_{0})$
are linearly dependent.  Analogously, since $s_{11}(z_{0})=0$ ($z_{0}\in \Upsilon\cap D_{-}$),  $\Phi_{+1}(x, t; z_{0})$ and $\Phi_{-2}(x, t; z_{0})$
can be linearly dependent. Then, we have
\begin{align}\label{68}
\Phi_{+2}(x, t, z_{0})=b[z_{0}]\Phi_{-1}(x, t, z_{0}),\quad z_{0}\in \Upsilon\cap D_{+},\notag\\
\Phi_{+1}(x, t, z_{0})=b[z_{0}]\Phi_{-2}(x, t, z_{0}),\quad z_{0}\in \Upsilon\cap D_{-},
\end{align}
where $b[z_{0}]$ is a norming constant. Due to $s'_{22}(z_{0})=0$ ($z_{0}\in \Upsilon\cap D_{+}$), we find that $\Phi'_{+2}(x, t; z_{0})-b[z_{0}]\Phi'_{-1}(x, t; z_{0})$ and $\Phi_{-1}(x, t; z_{0})$ are linearly
dependent.  Similarly, when $z_{0}\in \Upsilon\cap D_{-}$, $\Phi'_{+1}(x, t; z_{0})-b[z_{0}]\Phi'_{-2}(x, t; z_{0})$ and $\Phi_{-2}(x, t; z_{0})$ are linearly dependent. Then, we have
\begin{align}\label{69}
\Phi'_{+2}(x, t; z_{0})-b[z_{0}]\Phi'_{-1}(x, t; z_{0})=d[z_{0}]\Phi_{-1}(x, t; z_{0}),\quad z_{0}\in \Upsilon\cap D_{+},\notag\\
\Phi'_{+1}(x, t; z_{0})-b[z_{0}]\Phi'_{-2}(x, t; z_{0})=d[z_{0}]\Phi_{-2}(x, t; z_{0}),\quad z_{0}\in \Upsilon\cap D_{-},
\end{align}
where $d[z_{0}]$ is a norming constant. Therefore, one obtains
\begin{align}\label{70}
&\mathop{L_{-2}}_{z=z_{0}}\left[\frac{\Phi_{+2}(x, t; z)}{s_{22}(z)}\right]=A[z_{0}]\Phi_{-1}(x, t; z_{0}), \quad z_{0}\in \Upsilon\cap D_{+},\notag\\
&\mathop{L_{-2}}_{z=z_{0}}\left[\frac{\Phi_{+1}(x, t; z)}{s_{11}(z)}\right]=A[z_{0}]\Phi_{-2}(x, t; z_{0}), \quad z_{0}\in \Upsilon\cap D_{-},\notag\\
&\mathop{\mbox{Res}}_{z=z_{0}}\left[\frac{\Phi_{+2}(x, t; z)}{s_{22}(z)}\right]=A[z_{0}]\left[\Phi'_{-1}(x, t; z_{0})+B[z_{0}]\Phi_{-1}(x, t; z_{0})\right], \quad z_{0}\in \Upsilon\cap D_{+},\notag\\
&\mathop{\mbox{Res}}_{z=z_{0}}\left[\frac{\Phi_{+1}(x, t; z)}{s_{11}(z)}\right]=A[z_{0}]\left[\Phi'_{-2}(x, t; z_{0})+B[z_{0}]\Phi_{-2}(x, t; z_{0})\right], \quad z_{0}\in \Upsilon\cap D_{-},
\end{align}
where $L_{-2}[f(x, t; z)]$ is the coefficient of $O((z-z_{0})^{-2})$ term in the Laurent series expansion of $f(x, t; z)$ at $z=z_{0}$.
\begin{gather}
A[z_{0}]=\left\{
\begin{array}{lr}
\frac{2b[z_{0}]}{s''_{22}(z_{0})}, \quad z_{0}\in \Upsilon\cap D_{+}\\
\frac{2b[z_{0}]}{s''_{11}(z_{0})}, \quad z_{0}\in \Upsilon\cap D_{-}.
\end{array}
\right.\notag\\
B[z_{0}]=\left\{
\begin{array}{lr}
\frac{d[z_{0}]}{b[z_{0}]}-\frac{s'''_{22}(z_{0})}{3s''_{22}(z_{0})}, \quad z_{0}\in \Upsilon\cap D_{+}\\
\frac{d[z_{0}]}{b[z_{0}]}-\frac{s'''_{11}(z_{0})}{3s''_{11}(z_{0})}, \quad z_{0}\in \Upsilon\cap D_{-}.
\end{array}
\right. \label{71}
\end{gather}

\noindent \textbf{Proposition 6.1.} \emph{ Let $k_{0}\in \Upsilon$, then
the following symmetry relations are satisfied:
}\\
\emph{$\bullet$ The first symmetry relation $A[z_{0}]=-A[z^{\ast}_{0}]^{\ast}, B[z_{0}]=B[z^{\ast}_{0}]^{\ast}$;}\\
\emph{$\bullet$ The second symmetry relation $A[z_{0}]=A[-z^{\ast}_{0}]^{\ast}, B[z_{0}]=-B[-z^{\ast}_{0}]^{\ast}$;}\\
\emph{$\bullet$ The Third symmetry relation $A[z_{0}]=\frac{z_{0}^{4}u_{-}}{u_{0}^{4}u_{-}^{\ast}}A[-\frac{u_{0}^{2}}{z_{0}}]$,
$B[z_{0}]=\frac{u_{0}^{2}}{z_{0}^{2}}B[-\frac{u_{0}^{2}}{z_{0}}]+\frac{2}{z_{0}}$.}

As a matter of convenience, we set $\hat{\zeta}_{n}=-\frac{u_{0}^{2}}{\zeta_{n}}$
and
\begin{align}\label{77}
\zeta_{n}=\left\{
\begin{array}{lr}
z_{n}, \qquad n=1, 2, \cdots, N_{1}\\
-z_{n-N_{1}}, \qquad n=N_{1}+1, N_{1}+2, \cdots, 2N_{1}\\
\frac{u_{0}^{2}}{z_{n-2N_{1}}^{\ast}}, \qquad n=2N_{1}+1, 2N_{1}+2, \cdots, 3N_{1}\\
-\frac{u_{0}^{2}}{z_{n-3N_{1}}^{\ast}}, \qquad n=3N_{1}+1, 3N_{1}+2, \cdots, 4N_{1}\\
\omega_{n-4N_{1}}, \qquad n=4N_{1}+1, 4N_{1}+2, \cdots, 4N_{1}+N_{2}\\
-\omega_{n-4N_{1}-N_{2}}, \qquad n=4N_{1}+N_{2}+1, 4N_{1}+N_{2}+2, \cdots, 4N_{1}+2N_{2}\\
\end{array}
\right.
\end{align}
Then the residue and the coefficient $L_{-2}$ of $M(x, t; z)$ can be written as
\begin{align}\label{79}
&\mathop{L_{-2}}_{z=\zeta_{n}}M_{+}=\left(0, A[\zeta_{n}]e^{-2i\theta(x, t; \zeta_{n})}\mu_{-1}(x, t; \zeta_{n})\right),\notag\\
&\mathop{L_{-2}}_{z=\hat{\zeta}_{n}}M_{-}=\left(A[\hat{\zeta}_{n}]e^{2i\theta(x, t; \hat{\zeta}_{n})}\mu_{-2}(x, t; \hat{\zeta}_{n}), 0\right),\notag\\
&\mathop{\mbox{Res}}_{z=\zeta_{n}}M_{+}=\left(0, A[\zeta_{n}]e^{-2i\theta(x, t; \zeta_{n})}\left[\mu'_{-1}(x, t; \zeta_{n})+\left[B[\zeta_{n}]-2i\theta'(x, t; \zeta_{n})\right]\mu_{-1}(x, t; \zeta_{n})\right]\right),\notag\\
&\mathop{\mbox{Res}}_{z=\hat{\zeta}_{n}}M_{-}=\left(A[\hat{\zeta}_{n}]e^{2i\theta(x, t; \hat{\zeta}_{n})}\left[\mu'_{-2}(x, t; \hat{\zeta}_{n})+\left[B[\hat{\zeta}_{n}]+2i\theta'(x, t; \hat{\zeta}_{n})\right]\mu_{-2}(x, t; \hat{\zeta}_{n})\right], 0\right).
\end{align}

By subtracting out the asymptotic values as $z\rightarrow\infty, z\rightarrow 0$, the residue, and the coefficient $L_{-2}$ from the original non-regular RHP, one can obtain the following regular RHP
\begin{gather}
M_{-}+\frac{i}{z}\sigma_{3}Q_{-}-I-\notag\\
\sum_{n=1}^{4N_{1}+2N_{2}}\left[\frac{\mathop{L_{-2}}\limits_{z=\zeta_{n}}M_{+}}{(z-\zeta_{n})^{2}}
+\frac{\mathop{\mbox{Res}}\limits_{z=\zeta_{n}}M_{+}}{z-\zeta_{n}}+\frac{\mathop{L_{-2}}\limits_{z=\hat{\zeta}_{n}}M_{-}}
{(z-\hat{\zeta}_{n})^{2}}+\frac{\mathop{\mbox{Res}}\limits_{z=\hat{\zeta}_{n}}M_{-}}{(z-\hat{\zeta}_{n})}\right]=\notag\\
M_{+}+\frac{i}{z}\sigma_{3}Q_{-}-I-\notag\\
\sum_{n=1}^{4N_{1}+2N_{2}}\left[\frac{\mathop{L_{-2}}\limits_{z=\zeta_{n}}M_{+}}{(z-\zeta_{n})^{2}}
+\frac{\mathop{\mbox{Res}}\limits_{z=\zeta_{n}}M_{+}}{z-\zeta_{n}}+\frac{\mathop{L_{-2}}\limits_{z=\hat{\zeta}_{n}}M_{-}}
{(z-\hat{\zeta}_{n})^{2}}+\frac{\mathop{\mbox{Res}}\limits_{z=\hat{\zeta}_{n}}M_{-}}{(z-\hat{\zeta}_{n})}\right]-M_{+}G.\label{80}
\end{gather}
Via the Plemelj's formulae, the above RHP can be solved as
\begin{gather}
M(x, t; z)=I-\frac{i}{z}\sigma_{3}Q_{-}+\frac{1}{2\pi i}\int_{\Sigma}\frac{M_{+}(x, t; \zeta)G(x, t; \zeta)}{\zeta-z}d\zeta\notag\\
+\sum_{n=1}^{4N_{1}+2N_{2}}\left[\frac{\mathop{L_{-2}}\limits_{z=\zeta_{n}}M_{+}}{(z-\zeta_{n})^{2}}
+\frac{\mathop{\mbox{Res}}\limits_{z=\zeta_{n}}M_{+}}{z-\zeta_{n}}+\frac{\mathop{L_{-2}}\limits_{z=\hat{\zeta}_{n}}M_{-}}
{(z-\hat{\zeta}_{n})^{2}}+\frac{\mathop{\mbox{Res}}\limits_{z=\hat{\zeta}_{n}}M_{-}}{(z-\hat{\zeta}_{n})}\right].\label{81}
\end{gather}
where
\begin{gather}
\frac{\mathop{L_{-2}}\limits_{z=\zeta_{n}}M_{+}}{(z-\zeta_{n})^{2}}
+\frac{\mathop{\mbox{Res}}\limits_{z=\zeta_{n}}M_{+}}{z-\zeta_{n}}+\frac{\mathop{L_{-2}}\limits_{z=\hat{\zeta}_{n}}M_{-}}
{(z-\hat{\zeta}_{n})^{2}}+\frac{\mathop{\mbox{Res}}\limits_{z=\hat{\zeta}_{n}}M_{-}}{(z-\hat{\zeta}_{n})}=\notag\\
\left(\hat{C}_{n}(z)\left[\mu'_{-2}(\hat{\zeta}_{n})+\left(\hat{D}_{n}+\frac{1}{z-\hat{\zeta}_{n}}\right)\mu_{-2}(\hat{\zeta}_{n})\right],
C_{n}(z)\left[\mu'_{-1}(\zeta_{n})+\left(D_{n}+\frac{1}{z-\zeta_{n}}\right)\mu_{-1}(\zeta_{n})\right]\right),\label{82}
\end{gather}
and
\begin{align}\label{83}
&C_{n}(z)=\frac{A[\zeta_{n}]e^{-2i\theta(\zeta_{n})}}{z-\zeta_{n}}, \quad D_{n}=B[\zeta_{n}]-2i\theta'(\zeta_{n}), \notag\\
&\hat{C}_{n}(z)=\frac{A[\hat{\zeta}_{n}]e^{2i\theta(\hat{\zeta}_{n})}}{z-\hat{\zeta}_{n}}, \quad \hat{D}_{n}=B[\hat{\zeta}_{n}]+2i\theta'(\hat{\zeta}_{n}).
\end{align}

Moreover, according to \eqref{76}, we obtain
\begin{gather}\label{85}
M^{(1)}(x, t; z)=-\frac{1}{2\pi i}\int_{\Sigma}M_{+}(x, t; \zeta)G(x, t; \zeta)d\zeta -i\sigma_{3}Q_{-} \notag\\
+\sum_{n=1}^{4N_{1}+2N_{2}}\left(A[\hat{\zeta}_{n}]e^{2i\theta(\hat{\zeta}_{n})}\left(\mu'_{-2}(\hat{\zeta}_{n})+\hat{D}_{n}\mu_{-2}(\hat{\zeta}_{n})\right),
A[\zeta_{n}]e^{-2i\theta(\zeta_{n})}\left(\mu'_{-1}(\zeta_{n})+D_{n}\mu_{-1}(\zeta_{n})\right)\right).
\end{gather}
The potential $u(x, t)$ with double poles for the GI equation with NZBCs is given by
\begin{gather}
u(x, t)=iM_{12}^{(1)}=u_{-}
-\frac{1}{2\pi}\int_{\Sigma}(M_{+}(x, t; \zeta)G(x, t; \zeta))_{12}d\zeta \notag\\
+i\sum_{n=1}^{4N_{1}+2N_{2}}
A[\zeta_{n}]e^{-2i\theta(\zeta_{n})}\left(\mu'_{-11}(\zeta_{n})+D_{n}\mu_{-11}(\zeta_{n})\right).\label{86}
\end{gather}
\subsection{Trace formulae and theta condition}
Since $\zeta_{n}$ and $\hat{\zeta}_{n}$ are double zeros of the scattering coefficients $s_{22}(z)$ and $s_{11}(z)$, respectively, we can set
\begin{align}\label{jia1}
\beta^{+}(z)=s_{22}(z)\prod_{n=1}^{4N_{1}+2N_{2}}(\frac{z-\hat{\zeta}_{n}}{z-\zeta_{n}})^{2},\ \beta^{-}(z)=s_{22}(z)\prod_{n=1}^{4N_{1}+2N_{2}}(\frac{z-\zeta_{n}}{z-\hat{\zeta}_{n}})^{2},
\end{align}
such that $\beta^{+}(z)$ is analytic and has no zeros in $D_{+}$, and $\beta^{-}(z)$ is analytic and has no zeros in $D_{-}$.
They both tend to $o(1)$ as $z\rightarrow\infty$. Through using Cauchy projectors and the Plemelj's formulae, $\beta^{\pm}(z)$ can be expressed as
\begin{align}\label{jia2}
\log\beta^{\pm}(z)=\mp\frac{1}{2\pi i}\int_{\Sigma}\frac{\log(1-\rho\tilde{\rho})}{s-z}ds,\quad z\in D^{\pm}.
\end{align}
According to Eq.\eqref{jia1}, the trace formulae is given as
\begin{align}\label{jia3}
&s_{22}(z)=\mbox{exp}\left[-\frac{1}{2\pi i}\int_{\Sigma}\frac{\log(1-\rho\tilde{\rho})}{s-z}ds\right] \prod_{n=1}^{4N_{1}+2N_{2}}(\frac{z-\zeta_{n}}{z-\hat{\zeta}_{n}})^{2},\notag\\
&s_{11}(z)=\mbox{exp}\left[\frac{1}{2\pi i}\int_{\Sigma}\frac{\log(1-\rho\tilde{\rho})}{s-z}ds\right] \prod_{n=1}^{4N_{1}+2N_{2}}(\frac{z-\hat{\zeta}_{n}}{z-\zeta_{n}})^{2}.
\end{align}
Furthermore, let $z\rightarrow0$, one has
\begin{align}\label{jia4}
&\frac{u_{+}}{u_{-}}=\mbox{exp}\left[\frac{i}{2\pi}\int_{\Sigma}\frac{\log(1-\rho\tilde{\rho})}{s}ds\right] \prod_{n=1}^{N_{1}}(\frac{z_{n}}{z^{\ast}_{n}})^{8}\prod_{n=1}^{N_{2}}(\frac{\omega_{n}}{u_{0}})^{8}.
\end{align}
Therefore, we easily obtain the theta condition for Eq.\eqref{jia4}, given by
\begin{align}\label{jia5}
&\mbox{arg}(\frac{u_{+}}{u_{-}})=\frac{1}{2\pi}\int_{\Sigma}\frac{\log(1-\rho\tilde{\rho})}{s}ds +16\sum_{n=1}^{N_{1}}\mbox{arg}(z_{n})+8\sum_{n=1}^{N_{2}}\mbox{arg}(\omega_{n}).
\end{align}

\subsection{Double poles soliton solutions with NZBCs}
To get double poles soliton solution, we let $\rho(z)=\tilde{\rho}(z)=0$. Then, the second column of Eq.\eqref{81} yields
\begin{align}\label{87}
\mu_{-2}(z)=\left(\begin{array}{c}
    -\frac{i}{z}u_{-}\\
  1\\
\end{array}\right)+\sum_{n=1}^{4N_{1}+2N_{2}}C_{n}(z)\left[\mu'_{-1}(\zeta_{n})+\left(D_{n}+\frac{1}{z-\zeta_{n}}\right)\mu_{-1}(\zeta_{n})\right],
\end{align}
\begin{align}\label{88}
\mu_{-2}'(z)=\left(\begin{array}{c}
    \frac{i}{z^{2}}u_{-}\\
  0\\
\end{array}\right)-\sum_{n=1}^{4N_{1}+2N_{2}}\frac{C_{n}(z)}{z-\zeta_{n}}\left[\mu'_{-1}(\zeta_{n})+\left(D_{n}+\frac{2}{z-\zeta_{n}}\right)\mu_{-1}(\zeta_{n})\right].
\end{align}
According to the symmetric relation, one can obtain
\begin{align}\label{89}
\mu_{-2}(z)=-\frac{iu_{-}}{z}\mu_{-1}(-\frac{u_{0}^{2}}{z}).
\end{align}
Taking the first-order derivative about $z$ in above formula, we get
\begin{align}\label{90}
\mu_{-2}'(z)=\frac{iu_{-}}{z^{2}}\mu_{-1}(-\frac{u_{0}^{2}}{z})-\frac{iu_{-}u_{0}^{2}}{z^{3}}\mu'_{-1}(-\frac{u_{0}^{2}}{z}).
\end{align}
Putting Eqs.\eqref{89} and \eqref{90} into Eqs. \eqref{87} and \eqref{88},  and letting $z=\hat{\zeta}_{j}, j=1, 2,\cdots, 4N_{1}+
2N_{2}$, we obtain a $8N_{1}+4N_{2}$ linear system, given by
\begin{align}\label{91}
\sum_{n=1}^{4N_{1}+2N_{2}}\left\{C_{n}(\hat{\zeta}_{j})\mu'_{-1}(\zeta_{n})+\left[C_{n}(\hat{\zeta}_{j})\left(D_{n}+\frac{1}{\hat{\zeta}_{j}-\zeta_{n}}\right)+\frac{iu_{-}}{\hat{\zeta}_{j}}\delta_{j,n}\right]\mu_{-1}(\zeta_{n})
\right\}=\left(\begin{array}{c}
    \frac{i}{\hat{\zeta}_{j}}u_{-}\\
  -1\\
\end{array}\right),
\end{align}
\begin{gather}
\sum_{n=1}^{4N_{1}+2N_{2}}\left\{\left(\frac{C_{n}(\hat{\zeta}_{j})}{\hat{\zeta}_{j}-\zeta_{n}}-\frac{iu_{-}u_{0}^{2}}{\hat{\zeta}_{j}^{3}}\delta_{j,n}\right)
\mu'_{-1}(\zeta_{n})\right.\notag\\
\left.+\left[\frac{C_{n}(\hat{\zeta}_{j})}{\hat{\zeta}_{j}-\zeta_{n}}\left(D_{n}+\frac{2}{\hat{\zeta}_{j}-\zeta_{n}}\right)
+\frac{iu_{-}}{\hat{\zeta}_{j}^{2}}\delta_{j,n}\right]\mu_{-1}(\zeta_{n})\right\}
=\left(\begin{array}{c}
    \frac{i}{\hat{\zeta}_{j}^{2}}u_{-}\\
  0\\
\end{array}\right).\label{92}
\end{gather}

\noindent \textbf{Theorem 6.1}  \emph{
The general formula of the double poles solution for the GI equation \eqref{1} with NZBCs \eqref{49} is expressed as
\begin{align}\label{93}
u(x,t)=u_{-}-i\frac{\det \left(\begin{array}{cc}
    \mathcal{H}  &  \varphi\\
  \chi^{T} &  0\\
\end{array}\right)}{\det (\mathcal{H})},
\end{align}
where $\varphi, \mathcal{H}, \chi$ are given by \eqref{97}, \eqref{98}, \eqref{100}, respectively.
}
\begin{proof}
From Eqs.\eqref{91} and \eqref{92}, we get a $8N_{1}+4N_{2}$ linear system with respect to $\mu_{-11}(\zeta_{n})$, $\mu'_{-11}(\zeta_{n})$
\begin{align}\label{94}
\sum_{n=1}^{4N_{1}+2N_{2}}\left\{C_{n}(\hat{\zeta}_{j})\mu'_{-11}(\zeta_{n})+\left[C_{n}(\hat{\zeta}_{j})\left(D_{n}+\frac{1}{\hat{\zeta}_{j}-\zeta_{n}}\right)+\frac{iu_{-}}{\hat{\zeta}_{j}}\delta_{j,n}\right]\mu_{-11}(\zeta_{n})
\right\}=\frac{iu_{-}}{\hat{\zeta}_{j}}
\end{align}
\begin{gather}
\sum_{n=1}^{4N_{1}+2N_{2}}\left\{\left(\frac{C_{n}(\hat{\zeta}_{j})}{\hat{\zeta}_{j}-\zeta_{n}}-\frac{iu_{-}u_{0}^{2}}{\hat{\zeta}_{j}^{3}}\delta_{j,n}\right)
\mu'_{-11}(\zeta_{n})\right.\notag\\
\left.+\left[\frac{C_{n}(\hat{\zeta}_{j})}{\hat{\zeta}_{j}-\zeta_{n}}\left(D_{n}+\frac{2}{\hat{\zeta}_{j}-\zeta_{n}}\right)
+\frac{iu_{-}}{\hat{\zeta}_{j}^{2}}\delta_{j,n}\right]\mu_{-11}(\zeta_{n})\right\}
=\frac{iu_{-}}{\hat{\zeta}_{j}^{2}},\label{95}
\end{gather}
which can be rewritten in the matrix form:
\begin{align}\label{96}
\mathcal{H}\gamma=\varphi,
\end{align}
where
\begin{align}\label{97}
&\varphi=\left(\begin{array}{c}
     \varphi^{(1)}\\
  \varphi^{(2)}\\
\end{array}\right),\ \varphi^{(1)}=(\frac{iu_{-}}{\hat{\zeta}_{1}}, \frac{iu_{-}}{\hat{\zeta}_{2}},\cdots, \frac{iu_{-}}{\hat{\zeta}_{4N_{1}+2N_{2}}})^{T}, \varphi^{(2)}=(\frac{iu_{-}}{\hat{\zeta}_{1}^{2}}, \frac{iu_{-}}{\hat{\zeta}_{2}^{2}},\cdots, \frac{iu_{-}}{\hat{\zeta}_{4N_{1}+2N_{2}}^{2}})^{T}, \notag\\
&\gamma=\left(\begin{array}{c}
     \gamma^{(1)}\\
  \gamma^{(2)}\\
\end{array}\right),\ \gamma^{(1)}=(\mu_{-11}(\zeta_{1}), \mu_{-11}(\zeta_{2}),\cdots, \mu_{-11}(\zeta_{4N_{1}+2N_{2}}))^{T}, \notag\\
&\gamma^{(2)}=(\mu'_{-11}(\zeta_{1}), \mu'_{-11}(\zeta_{2}),\cdots, \mu'_{-11}(\zeta_{4N_{1}+2N_{2}}))^{T}, \mathcal{H}=\left(\begin{array}{cc}
    \mathcal{H}^{(11)} &  \mathcal{H}^{(12)}\\
  \mathcal{H}^{(21)} &  \mathcal{H}^{(22)}\\
\end{array}\right),
\end{align}
with $\mathcal{H}^{(im)}=\left(\mathcal{H}^{(im)}_{jn}\right)_{4N_{1}+2N_{2}\times 4N_{1}+2N_{2}}(i, m=
1, 2)$ given by
\begin{align}\label{98}
&\mathcal{H}^{(11)}_{jn}=C_{n}(\hat{\zeta}_{j})\left(D_{n}+\frac{1}{\hat{\zeta}_{j}-\zeta_{n}}\right)+\frac{iu_{-}}{\hat{\zeta}_{j}}\delta_{j,n}, \quad
\mathcal{H}^{(12)}_{jn}=C_{n}(\hat{\zeta}_{j}), \notag\\
&\mathcal{H}^{(21)}_{jn}=\frac{C_{n}(\hat{\zeta}_{j})}{\hat{\zeta}_{j}-\zeta_{n}}\left(D_{n}+\frac{2}{\hat{\zeta}_{j}-\zeta_{n}}\right)
+\frac{iu_{-}}{\hat{\zeta}_{j}^{2}}\delta_{j,n}, \quad
\mathcal{H}^{(22)}_{jn}=\frac{C_{n}(\hat{\zeta}_{j})}{\hat{\zeta}_{j}-\zeta_{n}}-\frac{iu_{-}u_{0}^{2}}{\hat{\zeta}_{j}^{3}}\delta_{j,n}.
\end{align}
According to the reflectionless potential, Eq.\eqref{86} can be rewritten as
\begin{align}\label{99}
u=u_{-}+i\chi^{T} \gamma,
\end{align}
where
\begin{align}\label{100}
&\chi=\left(\begin{array}{c}
     \chi^{(1)}\\
  \chi^{(2)}\\
\end{array}\right),
\chi^{(2)}=(A[\zeta_{1}]e^{-2i\theta(\zeta_{1})}, A[\zeta_{2}]e^{-2i\theta(\zeta_{2})},\cdots, A[\zeta_{4N_{1}+2N_{2}}]e^{-2i\theta(\zeta_{4N_{1}+2N_{2}})})^{T},\notag\\
&\chi^{(1)}=(A[\zeta_{1}]e^{-2i\theta(\zeta_{1})}D_{1}, A[\zeta_{2}]e^{-2i\theta(\zeta_{2})}D_{2},\cdots, A[\zeta_{4N_{1}+2N_{2}}]e^{-2i\theta(\zeta_{4N_{1}+2N_{2}})}D_{4N_{1}+2N_{2}})^{T}.
\end{align}
From Eqs. \eqref{96}, the expression of the double poles soliton solution is derived.
\end{proof}

As a example, through choosing some appropriate parameters, we discuss the dynamical behaviors for one-double and two-double poles soliton solution in the case of $N_{1}=0, N_{2}=1$,  $N_{1}=1, N_{2}=0$, and $N_{1}=1, N_{2}=1$, respectively. For $N_{1}=0, N_{2}=1$, we choose the parameters  $u_{-}=1, A[\omega_{1}]=1, B[\omega_{1}]=1, \omega_{1}=e^{\frac{\pi}{4}i}$ in which the theta condition is $\mbox{arg}(\frac{u_{+}}{u_{-}})=2\pi$ such that $u_{+}=u_{-}$. For this case,  the one-double poles soliton solution exhibits the interaction of dark and bright solitons, which can be verified in  Fig.6. Additionally, from Fig.6(c), it is easily to find that the collision is an elastic collision, due to the shape and size of the dark and bright solitons remain unchanged after the collision. On the other hand,  when $N_{1}=1, N_{2}=0$, we select the parameters $u_{-}=1, A[z_{1}]=1, B[z_{1}]=i, z_{1}=2e^{\frac{\pi}{6}i}$ and the theta condition is $\mbox{arg}(\frac{u_{+}}{u_{-}})=\frac{8}{3}\pi$ which leads to $u_{+}=e^{\frac{2}{3}\pi i}$. As shown in Fig.7, the one-double poles soliton solution stands for the interaction of two breather waves. Furthermore, when $N_{1}=N_{2}=1$ i.e., a mixed discrete spectra,  the two-breather-two-soliton solutions can be obtained, which displays the interaction of two-breather and two-soliton.\\
{\rotatebox{0}{\includegraphics[width=5.0cm,height=5.0cm,angle=0]{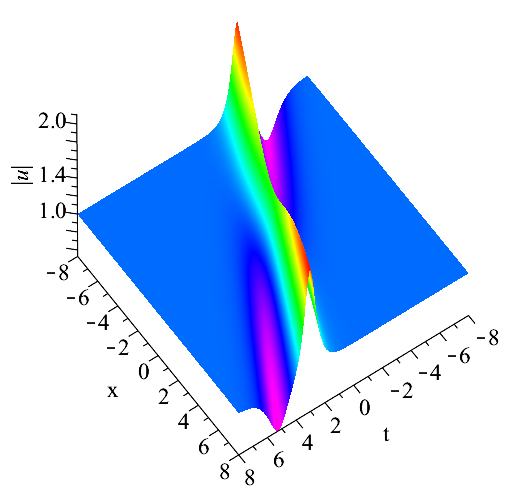}}}
~~~~\quad\qquad
{\rotatebox{0}{\includegraphics[width=5.0cm,height=5.0cm,angle=0]{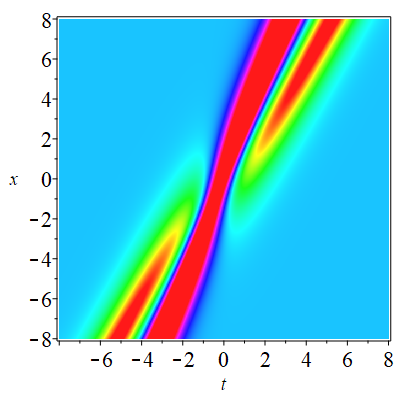}}}
~~~~\quad\qquad
{\rotatebox{0}{\includegraphics[width=5.0cm,height=5.0cm,angle=0]{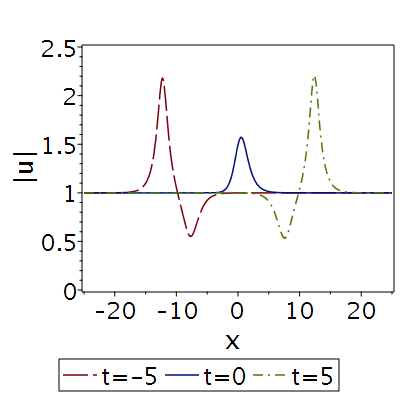}}}

$\qquad\qquad\textbf{(a)}\qquad\qquad\qquad\qquad\qquad\qquad\quad\qquad\qquad\quad 
\textbf{(b)}\qquad\qquad\qquad\qquad\qquad\qquad\qquad\qquad\quad\textbf{(c)}$\\
\noindent { \small \textbf{Figure 6.} (Color online) The one-double pole soliton solution for Eq.\eqref{1} with NZBCs and $N_{1}=0, N_{2}=1$. The parameters are $u_{\pm}=1, A[\omega_{1}]=1, B[\omega_{1}]=1, \omega_{1}=e^{\frac{\pi}{4}i}$. $\textbf{(a)}$ Three dimensional plot;
$\textbf{(b)}$ The density plot;
$\textbf{(c)}$ The wave propagation along the $x$-axis at $t=-5$(longdash), $t=0$(solid), $t=5$(dashdot).}\\
{\rotatebox{0}{\includegraphics[width=5.0cm,height=5.0cm,angle=0]{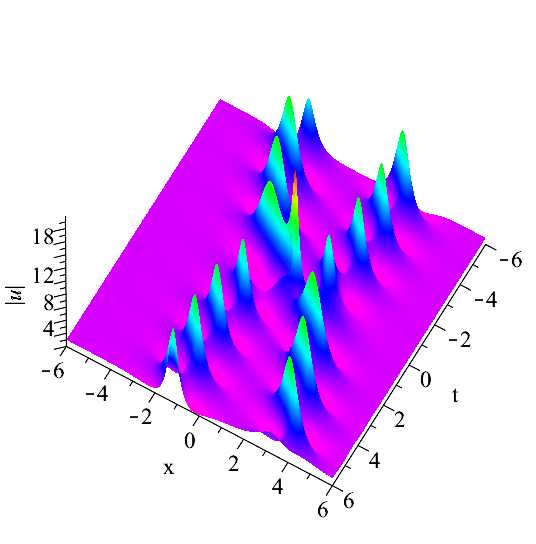}}}
~~~~\quad\qquad
{\rotatebox{0}{\includegraphics[width=5.0cm,height=5.0cm,angle=0]{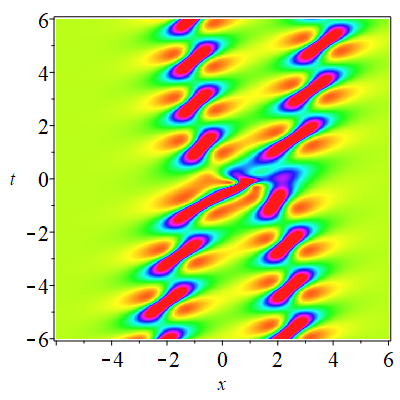}}}
~~~~\quad\qquad
{\rotatebox{0}{\includegraphics[width=5.0cm,height=5.0cm,angle=0]{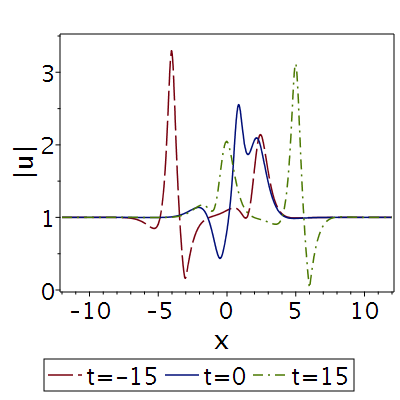}}}

$\qquad\qquad\textbf{(a)}\qquad\qquad\qquad\qquad\qquad\qquad\quad\qquad\qquad\quad 
\textbf{(b)}\qquad\qquad\qquad\qquad\qquad\qquad\qquad\qquad\quad\textbf{(c)}$\\
\noindent { \small \textbf{Figure 7.} (Color online) The one-double pole soliton solution for Eq.\eqref{1} with NZBCs and $N_{1}=1, N_{2}=0$. The parameters are $u_{-}=1, u_{+}=e^{\frac{2}{3}\pi i}, A[z_{1}]=1, B[z_{1}]=i, z_{1}=2e^{\frac{\pi}{6}i}$. $\textbf{(a)}$ Three dimensional plot;
$\textbf{(b)}$ The density plot;
$\textbf{(c)}$ The wave propagation along the $x$-axis at $t=-15$(longdash), $t=0$(solid), $t=15$(dashdot).}\\
%

\section{The solution of GI equation under NZBCs with triple poles}
In this section, we aim to derive the general $N$-triple poles solutions via analysing the inverse scattering problem with triple poles discrete spectrum for the GI equation \eqref{1} under NZBCs.
\subsection{Inverse scattering problem with NZBCs and triple poles}
Be similar to the expressions \eqref{68}, \eqref{69}, and since $s''_{22}(z_{0})=0$ in $z_{0}\in \Upsilon\cap D_{+}$ for the case of triple poles, it is not hard to find that $\Phi''_{+2}(x, t; z_{0})-b[z_{0}]\Phi''_{-1}(x, t; z_{0})-2d[z_{0}]\Phi'_{-1}(x, t; z_{0})$ and $\Phi_{-1}(x, t; z_{0})$ are linearly
dependent, and $\Phi''_{+1}(x, t; z_{0})-b[z_{0}]\Phi''_{-2}(x, t; z_{0})-2d[z_{0}]\Phi'_{-2}(x, t; z_{0})$ and $\Phi_{-2}(x, t; z_{0})$ are linearly dependent for $z_{0}\in \Upsilon\cap D_{-}$. That is to say
\begin{align}\label{21b}
\Phi''_{+2}(x, t; z_{0})-b[z_{0}]\Phi''_{-1}(x, t; z_{0})-2d[z_{0}]\Phi'_{-1}(x, t; z_{0})=h[z_{0}]\Phi_{-1}(x, t; z_{0}),\quad z_{0}\in \Upsilon\cap D_{+},\notag\\
\Phi''_{+1}(x, t; z_{0})-b[z_{0}]\Phi''_{-2}(x, t; z_{0})-2d[z_{0}]\Phi'_{-2}(x, t; z_{0})=h[z_{0}]\Phi_{-2}(x, t, z_{0}),\quad z_{0}\in \Upsilon\cap D_{-},
\end{align}
where $h[z_{0}]$ is also a norming constant.  Therefore, one has
\begin{align}\label{22b}
&\mathop{L_{-3}}_{z=z_{0}}\left[\frac{\Phi_{+2}(x, t; z)}{s_{22}(z)}\right]=\tilde{A}[z_{0}]\Phi_{-1}(x, t; z_{0}), \quad z_{0}\in \Upsilon\cap D_{+},\notag\\
&\mathop{L_{-3}}_{z=z_{0}}\left[\frac{\Phi_{+1}(x, t; z)}{s_{11}(z)}\right]=\tilde{A}[z_{0}]\Phi_{-2}(x, t; z_{0}), \quad z_{0}\in \Upsilon\cap D_{-},\notag\\
&\mathop{L_{-2}}_{z=z_{0}}\left[\frac{\Phi_{+2}(x, t; z)}{s_{22}(z)}\right]=\tilde{A}[z_{0}]\left[\Phi'_{-1}(x, t; z_{0})+\tilde{B}[z_{0}]\Phi_{-1}(x, t; z_{0})\right], \quad z_{0}\in \Upsilon\cap D_{+},\notag\\
&\mathop{L_{-2}}_{z=z_{0}}\left[\frac{\Phi_{+1}(x, t; z)}{s_{11}(z)}\right]=\tilde{A}[z_{0}]\left[\Phi'_{-2}(x, t; z_{0})+\tilde{B}[z_{0}]\Phi_{-2}(x, t; z_{0})\right], \quad z_{0}\in \Upsilon\cap D_{-},\notag\\
&\mathop{\mbox{Res}}_{z=z_{0}}\left[\frac{\Phi_{+2}(x, t; z)}{s_{22}(z)}\right]=\tilde{A}[z_{0}]\left[\frac{1}{2}\Phi''_{-1}(x, t; z_{0})+\tilde{B}[z_{0}]\Phi'_{-1}(x, t; z_{0})+\tilde{C}[z_{0}]\Phi_{-1}(x, t; z_{0})\right], \quad z_{0}\in \Upsilon\cap D_{+},\notag\\
&\mathop{\mbox{Res}}_{z=z_{0}}\left[\frac{\Phi_{+1}(x, t; z)}{s_{11}(z)}\right]=\tilde{A}[z_{0}]\left[\frac{1}{2}\Phi''_{-2}(x, t; z_{0})+\tilde{B}[z_{0}]\Phi'_{-2}(x, t; z_{0})+\tilde{C}[z_{0}]\Phi_{-2}(x, t; z_{0})\right], \quad z_{0}\in \Upsilon\cap D_{-},
\end{align}
where $L_{-3}[f(x, t; z)]$ means the coefficient of $O((z-z_{0})^{-3})$ term in the Laurent series expansion of $f(x, t; z)$ at $z=z_{0}$, and
\begin{align}\label{23b}
\tilde{A}[z_{0}]=\left\{
\begin{array}{lr}
\frac{6b[z_{0}]}{s'''_{22}(z_{0})}, \quad z_{0}\in \Upsilon\cap D_{+}\\
\\
\frac{6b[z_{0}]}{s'''_{11}(z_{0})}, \quad z_{0}\in \Upsilon\cap D_{-}.
\end{array}
\right.
\end{align}
\begin{align}\label{24b}
\tilde{B}[z_{0}]=\left\{
\begin{array}{lr}
\frac{d[z_{0}]}{b[z_{0}]}-\frac{s''''_{22}(z_{0})}{4s'''_{22}(z_{0})}, \quad z_{0}\in \Upsilon\cap D_{+}\\
\\
\frac{d[z_{0}]}{b[z_{0}]}-\frac{s''''_{11}(z_{0})}{4s'''_{11}(z_{0})}, \quad z_{0}\in \Upsilon\cap D_{-}.
\end{array}
\right.
\end{align}
\begin{align}\label{24b1}
\tilde{C}[z_{0}]=\left\{
\begin{array}{lr}
\frac{h[z_{0}]}{2b[z_{0}]}-\frac{d[z_{0}]s''''_{22}(z_{0})}{4b[z_{0}]s'''_{22}(z_{0})}+\frac{(s''''_{22})^{2}(z_{0})}{16(s'''_{22})^{2}(z_{0})}, \quad z_{0}\in \Upsilon\cap D_{+}\\
\\
\frac{h[z_{0}]}{2b[z_{0}]}-\frac{d[z_{0}]s''''_{11}(z_{0})}{4b[z_{0}]s'''_{11}(z_{0})}+\frac{(s''''_{11})^{2}(z_{0})}{16(s'''_{11})^{2}(z_{0})}, \quad z_{0}\in \Gamma\cap D_{-}.
\end{array}
\right.
\end{align}

\noindent \textbf{Proposition 7.1.} \emph{ Let $k_{0}\in \Upsilon$, then
the following symmetry relations are satisfied:
}\\
\emph{$\bullet$ The first symmetry relation $\tilde{A}[z_{0}]=-\tilde{A}[z^{\ast}_{0}]^{\ast}, \tilde{B}[z_{0}]=\tilde{B}[z^{\ast}_{0}]^{\ast}, \tilde{C}[z_{0}]=\tilde{C}[z^{\ast}_{0}]^{\ast}$;}\\
\emph{$\bullet$ The second symmetry relation $\tilde{A}[z_{0}]=-\tilde{A}[-z^{\ast}_{0}]^{\ast}, \tilde{B}[z_{0}]=-\tilde{B}[-z^{\ast}_{0}]^{\ast}, \tilde{C}[z_{0}]=\tilde{C}[-z^{\ast}_{0}]^{\ast}$;}\\
\emph{$\bullet$ The Third symmetry relation $\tilde{A}[z_{0}]=\frac{z_{0}^{6}u_{-}}{u_{0}^{6}u_{-}^{\ast}}\tilde{A}[-\frac{u_{0}^{2}}{z_{0}}], \tilde{B}[z_{0}]=\frac{u_{0}^{2}}{z_{0}^{2}}\tilde{B}[-\frac{u_{0}^{2}}{z_{0}}]+\frac{3}{z_{0}}, \tilde{C}[z_{0}]=\frac{u_{0}^{4}}{z_{0}^{4}}\tilde{C}[-\frac{u_{0}^{2}}{z_{0}}]
+\frac{2u_{0}^{3}}{z_{0}^{3}}\tilde{B}[-\frac{u_{0}^{2}}{z_{0}}]+\frac{3}{z_{0}^{2}}$.}

The residue and the coefficient $L_{-2}, L_{-3}$ of $M(x, t; z)$ are
\begin{align}\label{31b}
&\mathop{L_{-3}}_{z=\zeta_{n}}M_{+}=\left(0, \tilde{A}[\zeta_{n}]e^{-2i\theta(\zeta_{n})}\mu_{-1}(\zeta_{n})\right),\notag\\
&\mathop{L_{-3}}_{z=\hat{\zeta}_{n}}M_{-}=\left(\tilde{A}[\hat{\zeta}_{n}]e^{2i\theta(\hat{\zeta}_{n})}\mu_{-2}(\hat{\zeta}_{n}), 0\right),\notag\\
&\mathop{L_{-2}}_{z=\zeta_{n}}M_{+}=\left(0, \tilde{A}[\zeta_{n}]e^{-2i\theta(\zeta_{n})}\left[\mu'_{-1}(x, t; \zeta_{n})+\left[\tilde{B}[\zeta_{n}]-2i\theta'(\zeta_{n})\right]\mu_{-1}(\zeta_{n})\right]\right),\notag\\
&\mathop{L_{-2}}_{z=\hat{\zeta}_{n}}M_{-}=\left(\tilde{A}[\hat{\zeta}_{n}]e^{2i\theta(\hat{\zeta}_{n})}\left[\mu'_{-2}( \hat{\zeta}_{n})+\left[\tilde{B}[\hat{\zeta}_{n}]+2i\theta'(\hat{\zeta}_{n})\right]\mu_{-2}(\hat{\zeta}_{n})\right], 0\right),\notag\\
&\mathop{\mbox{Res}}_{z=\zeta_{n}}M_{+}=\left(0, \tilde{A}[\zeta_{n}]e^{-2i\theta(\zeta_{n})}\left[\frac{1}{2}\mu''_{-1}( \zeta_{n})+\left[\tilde{B}[\zeta_{n}]-2i\theta'(\zeta_{n})\right]\mu'_{-1}(\zeta_{n})+[\tilde{C}[\zeta_{n}]-\Theta_{1}(\zeta_{n})]\mu_{-1}( \zeta_{n})\right]\right),\notag\\
&\mathop{\mbox{Res}}_{z=\hat{\zeta}_{n}}M_{-}=\left(\tilde{A}[\hat{\zeta}_{n}]e^{2i\theta(\hat{\zeta}_{n})}\left[\frac{1}{2}\mu''_{-2}( \hat{\zeta}_{n})+\left[\tilde{B}[\hat{\zeta}_{n}]+2i\theta'(\hat{\zeta}_{n})\right]\mu'_{-2}( \hat{\zeta}_{n})+[\tilde{C}[\hat{\zeta}_{n}]-\Theta_{2}(\hat{\zeta}_{n})]\mu_{-2}(\hat{\zeta}_{n})\right], 0\right),
\end{align}
where
\begin{align}\label{31b1}
&\Theta_{1}(\zeta_{n})=2(\theta'(\zeta_{n}))^{2}+i\theta''(\zeta_{n})+2\tilde{B}[\zeta_{n}]i\theta'(\zeta_{n}),\notag\\
&\Theta_{2}(\hat{\zeta}_{n})=2(\theta'(\hat{\zeta}_{n}))^{2}-i\theta''(\hat{\zeta}_{n})-2\tilde{B}[\hat{\zeta}_{n}]i\theta'(\hat{\zeta}_{n}).
\end{align}
By subtracting out the residue, the coefficient $L_{-2}, L_{-3}$ and the asymptotic values as $z\rightarrow\infty$ from the original non-regular RHP, the following regular RHP is derived
\begin{gather}
M_{-}+\frac{i}{z}\sigma_{3}Q_{-}-I-\notag\\
\sum_{n=1}^{4N_{1}+2N_{2}}\left[\frac{\mathop{L_{-3}}\limits_{z=\zeta_{n}}M_{+}}{(z-\zeta_{n})^{3}}
+\frac{\mathop{L_{-2}}\limits_{z=\zeta_{n}}M_{+}}{(z-\zeta_{n})^{2}}
+\frac{\mathop{\mbox{Res}}\limits_{z=\zeta_{n}}M_{+}}{z-\zeta_{n}}+\frac{\mathop{L_{-3}}\limits_{z=\hat{\zeta}_{n}}M_{-}}
{(z-\hat{\zeta}_{n})^{3}}+\frac{\mathop{L_{-2}}\limits_{z=\hat{\zeta}_{n}}M_{-}}
{(z-\hat{\zeta}_{n})^{2}}+\frac{\mathop{\mbox{Res}}\limits_{z=\hat{\zeta}_{n}}M_{-}}{(z-\hat{\zeta}_{n})}\right]=\notag\\
M_{+}+\frac{i}{z}\sigma_{3}Q_{-}-I-\notag\\
\sum_{n=1}^{4N_{1}+2N_{2}}\left[\frac{\mathop{L_{-3}}\limits_{z=\zeta_{n}}M_{+}}{(z-\zeta_{n})^{3}}
+\frac{\mathop{L_{-2}}\limits_{z=\zeta_{n}}M_{+}}{(z-\zeta_{n})^{2}}
+\frac{\mathop{\mbox{Res}}\limits_{z=\zeta_{n}}M_{+}}{z-\zeta_{n}}+\frac{\mathop{L_{-3}}\limits_{z=\hat{\zeta}_{n}}M_{-}}
{(z-\hat{\zeta}_{n})^{3}}+\frac{\mathop{L_{-2}}\limits_{z=\hat{\zeta}_{n}}M_{-}}
{(z-\hat{\zeta}_{n})^{2}}+\frac{\mathop{\mbox{Res}}\limits_{z=\hat{\zeta}_{n}}M_{-}}{(z-\hat{\zeta}_{n})}\right]-M_{+}G,\label{32b}
\end{gather}
which can be solved by the Plemelj's formulae, given by
\begin{gather}
M(x, t; z)=I-\frac{i}{z}\sigma_{3}Q_{-}+\frac{1}{2\pi i}\int_{\Sigma}\frac{M_{+}(x, t; \zeta)G(x, t; \zeta)}{\zeta-z}d\zeta\notag\\
+\sum_{n=1}^{4N_{1}+2N_{2}}\left[\frac{\mathop{L_{-3}}\limits_{z=\zeta_{n}}M_{+}}{(z-\zeta_{n})^{3}}
+\frac{\mathop{L_{-2}}\limits_{z=\zeta_{n}}M_{+}}{(z-\zeta_{n})^{2}}
+\frac{\mathop{\mbox{Res}}\limits_{z=\zeta_{n}}M_{+}}{z-\zeta_{n}}+\frac{\mathop{L_{-3}}\limits_{z=\hat{\zeta}_{n}}M_{-}}
{(z-\hat{\zeta}_{n})^{3}}+\frac{\mathop{L_{-2}}\limits_{z=\hat{\zeta}_{n}}M_{-}}
{(z-\hat{\zeta}_{n})^{2}}+\frac{\mathop{\mbox{Res}}\limits_{z=\hat{\zeta}_{n}}M_{-}}{(z-\hat{\zeta}_{n})}\right],\label{33b}
\end{gather}
where
\begin{gather}
\frac{\mathop{L_{-3}}\limits_{z=\zeta_{n}}M_{+}}{(z-\zeta_{n})^{3}}
+\frac{\mathop{L_{-2}}\limits_{z=\zeta_{n}}M_{+}}{(z-\zeta_{n})^{2}}
+\frac{\mathop{\mbox{Res}}\limits_{z=\zeta_{n}}M_{+}}{z-\zeta_{n}}+\frac{\mathop{L_{-3}}\limits_{z=\hat{\zeta}_{n}}M_{-}}
{(z-\hat{\zeta}_{n})^{3}}+\frac{\mathop{L_{-2}}\limits_{z=\hat{\zeta}_{n}}M_{-}}
{(z-\hat{\zeta}_{n})^{2}}+\frac{\mathop{\mbox{Res}}\limits_{z=\hat{\zeta}_{n}}M_{-}}{z-\hat{\zeta}_{n}}=\notag\\
\left(\hat{C}_{n}(z)\left[\frac{1}{2}\mu''_{-2}(\hat{\zeta}_{n})+\left(\hat{D}_{n}+\frac{1}{z-\hat{\zeta}_{n}}\right)\mu'_{-2}(\hat{\zeta}_{n})+(\frac{1}{(z-\hat{\zeta}_{n})^{2}}+\frac{\hat{D}_{n}}{z-\hat{\zeta}_{n}}+\hat{F}_{n})\mu_{-2}(\hat{\zeta}_{n})\right],\right.\notag\\
\left.C_{n}(z)\left[\frac{1}{2}\mu''_{-1}(\zeta_{n})+\left(D_{n}+\frac{1}{z-\zeta_{n}}\right)\mu'_{-1}(\zeta_{n})+(\frac{1}{(z-\zeta_{n})^{2}}+\frac{D_{n}}{z-\zeta_{n}}+F_{n})\mu_{-1}(\zeta_{n})\right]\right),\label{34b}
\end{gather}
and
\begin{align}\label{35b}
&C_{n}(z)=\frac{\tilde{A}[\zeta_{n}]e^{-2i\theta(\zeta_{n})}}{z-\zeta_{n}}, \ D_{n}=\tilde{B}[\zeta_{n}]-2i\theta'(\zeta_{n}), \ F_{n}=\tilde{C}[\zeta_{n}]-\Theta_{1}(\zeta_{n}), \notag\\
&\hat{C}_{n}(z)=\frac{\tilde{A}[\hat{\zeta}_{n}]e^{2i\theta(\hat{\zeta}_{n})}}{z-\hat{\zeta}_{n}}, \ \hat{D}_{n}=\tilde{B}[\hat{\zeta}_{n}]+2i\theta'(\hat{\zeta}_{n}),\ \hat{F}_{n}=\tilde{C}[\hat{\zeta}_{n}]-\Theta_{2}(\hat{\zeta}_{n}).
\end{align}
Furthermore, according to \eqref{76}, we get
\begin{gather}
M^{(1)}(x, t; z)=-\frac{1}{2\pi i}\int_{\Sigma}M_{+}(x, t; \zeta)G(x, t; \zeta)d\zeta -i \sigma_{3}Q_{-}\notag\\
\sum_{n=1}^{4N_{1}+2N_{2}}\left(\tilde{A}[\hat{\zeta}_{n}]e^{2i\theta(\hat{\zeta}_{n})}\left(\frac{1}{2}\mu''_{-2}(\hat{\zeta}_{n})+\hat{D}_{n}\mu'_{-2}(\hat{\zeta}_{n})+\hat{F}_{n}\mu_{-2}(\hat{\zeta}_{n})\right),\right.\notag\\
\left.\tilde{A}[\zeta_{n}]e^{-2i\theta(\zeta_{n})}\left(\frac{1}{2}\mu''_{-1}(\zeta_{n})+D_{n}\mu'_{-1}(\zeta_{n})+F_{n}\mu_{1}(\zeta_{n})\right)\right). \label{37b}
\end{gather}
The potential $u(x, t)$ with triple poles for the GI equation with NZBCs turns into
\begin{gather}
u(x, t)=iM_{12}^{(1)}=u_{-}-\frac{1}{2\pi}\int_{\Sigma}(M_{+}(x, t; \zeta)G(x, t; \zeta))_{12}d\zeta \notag\\
+i\sum_{n=1}^{4N_{1}+2N_{2}}\tilde{A}[\zeta_{n}]e^{-2i\theta(\zeta_{n})}\left(\frac{1}{2}\mu''_{-11}(\zeta_{n})+D_{n}\mu'_{-11}(\zeta_{n})+F_{n}\mu_{11}(\zeta_{n})\right)
.\label{38b}
\end{gather}
\subsection{Trace formulae and theta condition}
In consideration of $\zeta_{n}$ and $\hat{\zeta}_{n}$ being trace zeros of the scattering coefficients $s_{22}(z)$ and $s_{11}(z)$, respectively, we introduce following functions
\begin{align}\label{jia6}
\beta^{+}(z)=s_{22}(z)\prod_{n=1}^{4N_{1}+2N_{2}}(\frac{z-\hat{\zeta}_{n}}{z-\zeta_{n}})^{3},\ \beta^{-}(z)=s_{22}(z)\prod_{n=1}^{4N_{1}+2N_{2}}(\frac{z-\zeta_{n}}{z-\hat{\zeta}_{n}})^{3},
\end{align}
of which $\beta^{+}(z)$ is analytic and has no zeros in $D_{+}$, and $\beta^{-}(z)$ is analytic and has no zeros in $D_{-}$. As $z\rightarrow\infty$,
they both tend to $o(1)$. Furthermore, $\beta^{\pm}(z)$ can be shown as
\begin{align}\label{jia7}
\log\beta^{\pm}(z)=\mp\frac{1}{2\pi i}\int_{\Sigma}\frac{\log(1-\rho\tilde{\rho})}{s-z}ds,\quad z\in D^{\pm},
\end{align}
and the trace formulae is
\begin{align}\label{jia8}
&s_{22}(z)=\mbox{exp}\left[-\frac{1}{2\pi i}\int_{\Sigma}\frac{\log(1-\rho\tilde{\rho})}{s-z}ds\right] \prod_{n=1}^{4N_{1}+2N_{2}}(\frac{z-\zeta_{n}}{z-\hat{\zeta}_{n}})^{3},\notag\\
&s_{11}(z)=\mbox{exp}\left[\frac{1}{2\pi i}\int_{\Sigma}\frac{\log(1-\rho\tilde{\rho})}{s-z}ds\right] \prod_{n=1}^{4N_{1}+2N_{2}}(\frac{z-\hat{\zeta}_{n}}{z-\zeta_{n}})^{3}.
\end{align}
Let $z\rightarrow0$, it arrives in
\begin{align}\label{jia9}
&\frac{u_{+}}{u_{-}}=\mbox{exp}\left[\frac{i}{2\pi}\int_{\Sigma}\frac{\log(1-\rho\tilde{\rho})}{s}ds\right] \prod_{n=1}^{N_{1}}(\frac{z_{n}}{z^{\ast}_{n}})^{12}\prod_{n=1}^{N_{2}}(\frac{\omega_{n}}{u_{0}})^{12}.
\end{align}
Ultimately, the theta condition for Eq.\eqref{jia9} is given by
\begin{align}\label{jia10}
&\mbox{arg}(\frac{u_{+}}{u_{-}})=\frac{1}{2\pi}\int_{\Sigma}\frac{\log(1-\rho\tilde{\rho})}{s}ds +24\sum_{n=1}^{N_{1}}\mbox{arg}(z_{n})+12\sum_{n=1}^{N_{2}}\mbox{arg}(\omega_{n}).
\end{align}

\subsection{Triple poles soliton solutions with NZBCs}
We take $\rho(z)=\tilde{\rho}(z)=0$ to generate the explicit triple poles soliton solutions of the GI equation with NZBCs. Then, the second column of Eq.\eqref{33b} yields
\begin{gather}
\mu_{-2}(z)=\left(\begin{array}{c}
    -\frac{i}{z}u_{-}\\
  1\\
\end{array}\right)+\notag\\
\sum_{n=1}^{4N_{1}+2N_{2}}C_{n}(z)\left[\frac{1}{2}\mu''_{-1}(\zeta_{n})
+\left(D_{n}+\frac{1}{z-\zeta_{n}}\right)\mu'_{-1}(\zeta_{n})+\left(\frac{1}{(z-\zeta_{n})^{2}}+\frac{D_{n}}{z-\zeta_{n}}+F_{n}\right)\mu_{-1}(\zeta_{n})\right],\notag\\
\mu_{-2}'(z)=\left(\begin{array}{c}
    \frac{i}{z^{2}}u_{-}\\
  0\\
\end{array}\right)-\notag\\
\sum_{n=1}^{4N_{1}+2N_{2}}\frac{C_{n}(z)}{z-\zeta_{n}}\left[\frac{1}{2}\mu''_{-1}(\zeta_{n})
+\left(D_{n}+\frac{2}{z-\zeta_{n}}\right)\mu'_{-1}(\zeta_{n})+\left(\frac{3}{(z-\zeta_{n})^{2}}+\frac{2D_{n}}{z-\zeta_{n}}+F_{n}\right)\mu_{-1}(\zeta_{n})\right]\notag\\
\mu_{-2}''(z)=\left(\begin{array}{c}
    \frac{-2i}{z^{3}}u_{-}\\
  0\\
\end{array}\right)+\notag\\
\sum_{n=1}^{2N}\frac{2C_{n}(z)}{(z-\xi_{n})^{2}}\left[\frac{1}{2}\mu''_{-1}(\zeta_{n})
+\left(D_{n}+\frac{3}{z-\zeta_{n}}\right)\mu'_{-1}(\zeta_{n})+\left(\frac{6}{(z-\zeta_{n})^{2}}+\frac{3D_{n}}{z-\zeta_{n}}+F_{n}\right)\mu_{-1}(\zeta_{n})\right].\label{39b}
\end{gather}
Taking the second-order derivative about $z$ in above formula \eqref{89}, we get
\begin{align}\label{90b}
\mu_{-2}''(z)=-\frac{2iu_{-}}{z^{3}}\mu_{-1}(-\frac{u_{0}^{2}}{z})
+\frac{4iu_{-}u_{0}^{2}}{z^{4}}\mu'_{-1}(-\frac{u_{0}^{2}}{z})
-\frac{iu_{-}u_{0}^{4}}{z^{5}}\mu''_{-1}(-\frac{u_{0}^{2}}{z}).
\end{align}
Putting Eqs.\eqref{89},\eqref{90} and \eqref{90b} into Eqs. \eqref{39b},  and letting $z=\hat{\zeta}_{j}, j=1, 2,\cdots, 4N_{1}+
2N_{2}$, we obtain a $12N_{1}+6N_{2}$ following linear system
\begin{gather}
\sum_{n=1}^{4N_{1}+2N_{2}}\left\{C_{n}(\hat{\zeta}_{j})\left(\frac{1}{2}\mu''_{-1}(\zeta_{n})
+\left(D_{n}+\frac{1}{\hat{\zeta}_{j}-\zeta_{n}}\right)\mu'_{-1}(\zeta_{n})\right)\right.\notag\\
\left.+\left[C_{n}(\hat{\zeta}_{j})\left(\frac{1}{(\hat{\zeta}_{j}-\zeta_{n})^{2}}+\frac{D_{n}}{\hat{\zeta}_{j}-\zeta_{n}}+F_{n}\right)
+\frac{iu_{-}}{\hat{\zeta}_{j}}\delta_{j,n}\right]\mu_{-1}(\zeta_{n})
\right\}=\left(\begin{array}{c}
    \frac{i}{\hat{\zeta}_{j}}u_{-}\\
  -1\\
\end{array}\right),\label{91b}
\end{gather}
\begin{gather}
\sum_{n=1}^{4N_{1}+2N_{2}}\left\{\frac{C_{n}(\hat{\zeta}_{j})}{2(\hat{\zeta}_{j}-\zeta_{n})}\mu''_{-1}(\zeta_{n})
+\left[\frac{C_{n}(\hat{\zeta}_{j})}{\hat{\zeta}_{j}-\zeta_{n}}\left(D_{n}+\frac{2}{\hat{\zeta}_{j}-\zeta_{n}}\right)-\frac{iu_{0}^{2}u_{-}}{\hat{\zeta}^{3}_{j}}\delta_{j,n}\right]\mu'_{-1}(\zeta_{n})\right.\notag\\
\left.+\left[\frac{C_{n}(\hat{\zeta}_{j})}{\hat{\zeta}_{j}-\zeta_{n}}\left(\frac{3}{(\hat{\zeta}_{j}-\zeta_{n})^{2}}+\frac{2D_{n}}{\hat{\zeta}_{j}-\zeta_{n}}+F_{n}\right)
+\frac{iu_{-}}{\hat{\zeta}^{2}_{j}}\delta_{j,n}\right]\mu_{-1}(\zeta_{n})
\right\}=\left(\begin{array}{c}
    \frac{i}{\hat{\zeta}^{2}_{j}}u_{-}\\
  0\\
\end{array}\right),\label{92b}
\end{gather}
\begin{gather}
\sum_{n=1}^{4N_{1}+2N_{2}}\left\{\left[\frac{C_{n}(\hat{\zeta}_{j})}{(\hat{\zeta}_{j}-\zeta_{n})^{2}}+\frac{iu_{0}^{4}u_{-}}{\hat{\zeta}^{5}_{j}}\delta_{j,n}\right]\mu''_{-1}(\zeta_{n})
+\left[\frac{2C_{n}(\hat{\zeta}_{j})}{(\hat{\zeta}_{j}-\zeta_{n})^{2}}\left(D_{n}+\frac{3}{\hat{\zeta}_{j}-\zeta_{n}}\right)-\frac{4iu_{0}^{2}u_{-}}{\hat{\zeta}^{4}_{j}}\delta_{j,n}\right]\mu'_{-1}(\zeta_{n})\right.\notag\\
\left.+\left[\frac{2C_{n}(\hat{\zeta}_{j})}{(\hat{\zeta}_{j}-\zeta_{n})^{2}}\left(\frac{6}{(\hat{\zeta}_{j}-\zeta_{n})^{2}}+\frac{3D_{n}}{\hat{\zeta}_{j}-\zeta_{n}}+F_{n}\right)
+\frac{2iu_{-}}{\hat{\zeta}^{3}_{j}}\delta_{j,n}\right]\mu_{-1}(\zeta_{n})
\right\}=\left(\begin{array}{c}
    \frac{2i}{\hat{\zeta}^{3}_{j}}u_{-}\\
  0\\
\end{array}\right),\label{92b1}
\end{gather}

\noindent \textbf{Theorem 7.1}  \emph{
The general formula of the triple poles solution for the GI equation \eqref{1} with NZBCs \eqref{49} is expressed as
\begin{align}\label{93b}
u(x,t)=u_{-}-i\frac{\det \left(\begin{array}{cc}
    \mathcal{\tilde{H}}  &  \tilde{\varphi}\\
  \tilde{\chi}^{T} &  0\\
\end{array}\right)}{\det (\mathcal{\tilde{H}})},
\end{align}
where $\tilde{\varphi}, \mathcal{\tilde{H}}, \tilde{\chi}$ are given by \eqref{97b}, \eqref{98b}, \eqref{100b}, respectively.
}
\begin{proof}
From Eq.\eqref{91b}, \eqref{92b} and \eqref{92b1}, we get a $12N_{1}+6N_{2}$ linear system with respect to $\mu_{-11}(\zeta_{n})$, $\mu'_{-11}(\zeta_{n})$
\begin{gather}
\sum_{n=1}^{4N_{1}+2N_{2}}\left\{C_{n}(\hat{\zeta}_{j})\left(\frac{1}{2}\mu''_{-11}(\zeta_{n})
+\left(D_{n}+\frac{1}{\hat{\zeta}_{j}-\zeta_{n}}\right)\mu'_{-11}(\zeta_{n})\right)\right.\notag\\
\left.+\left[C_{n}(\hat{\zeta}_{j})\left(\frac{1}{(\hat{\zeta}_{j}-\zeta_{n})^{2}}+\frac{D_{n}}{\hat{\zeta}_{j}-\zeta_{n}}+F_{n}\right)
+\frac{iu_{-}}{\hat{\zeta}_{j}}\delta_{j,n}\right]\mu_{-11}(\zeta_{n})
\right\}=\frac{i}{\hat{\zeta}_{j}}u_{-},\label{94b}
\end{gather}
\begin{gather}
\sum_{n=1}^{4N_{1}+2N_{2}}\left\{\frac{C_{n}(\hat{\zeta}_{j})}{2(\hat{\zeta}_{j}-\zeta_{n})}\mu''_{-11}(\zeta_{n})
+\left[\frac{C_{n}(\hat{\zeta}_{j})}{\hat{\zeta}_{j}-\zeta_{n}}\left(D_{n}+\frac{2}{\hat{\zeta}_{j}-\zeta_{n}}\right)-\frac{iu_{0}^{2}u_{-}}{\hat{\zeta}^{3}_{j}}\delta_{j,n}\right]\mu'_{-11}(\zeta_{n})\right.\notag\\
\left.+\left[\frac{C_{n}(\hat{\zeta}_{j})}{\hat{\zeta}_{j}-\zeta_{n}}\left(\frac{3}{(\hat{\zeta}_{j}-\zeta_{n})^{2}}+\frac{2D_{n}}{\hat{\zeta}_{j}-\zeta_{n}}+F_{n}\right)
+\frac{iu_{-}}{\hat{\zeta}^{2}_{j}}\delta_{j,n}\right]\mu_{-11}(\zeta_{n})
\right\}=\frac{i}{\hat{\zeta}^{2}_{j}}u_{-},\label{95b}
\end{gather}
\begin{gather}
\sum_{n=1}^{4N_{1}+2N_{2}}\left\{\left[\frac{C_{n}(\hat{\zeta}_{j})}{(\hat{\zeta}_{j}-\zeta_{n})^{2}}+\frac{iu_{0}^{4}u_{-}}{\hat{\zeta}^{5}_{j}}\delta_{j,n}\right]\mu''_{-11}(\zeta_{n})
+\left[\frac{2C_{n}(\hat{\zeta}_{j})}{(\hat{\zeta}_{j}-\zeta_{n})^{2}}\left(D_{n}+\frac{3}{\hat{\zeta}_{j}-\zeta_{n}}\right)-\frac{4iu_{0}^{2}u_{-}}{\hat{\zeta}^{4}_{j}}\delta_{j,n}\right]\mu'_{-11}(\zeta_{n})\right.\notag\\
\left.+\left[\frac{2C_{n}(\hat{\zeta}_{j})}{(\hat{\zeta}_{j}-\zeta_{n})^{2}}\left(\frac{6}{(\hat{\zeta}_{j}-\zeta_{n})^{2}}+\frac{3D_{n}}{\hat{\zeta}_{j}-\zeta_{n}}+F_{n}\right)
+\frac{2iu_{-}}{\hat{\zeta}^{3}_{j}}\delta_{j,n}\right]\mu_{-11}(\zeta_{n})
\right\}=\frac{2i}{\hat{\zeta}^{3}_{j}}u_{-},\label{95b1}
\end{gather}
which can be rewritten in the matrix form:
\begin{align}\label{96b}
\mathcal{\tilde{H}}\tilde{\gamma}=\tilde{\varphi},
\end{align}
where
\begin{gather}
\tilde{\varphi}=\left(\begin{array}{c}
     \tilde{\varphi}^{(1)}\\
  \tilde{\varphi}^{(2)}\\
   \tilde{\varphi}^{(3)}\\
\end{array}\right),\ \tilde{\varphi}^{(1)}=(\frac{iu_{-}}{\hat{\zeta}_{1}},\cdots, \frac{iu_{-}}{\hat{\zeta}_{4N_{1}+2N_{2}}})^{T}, \tilde{\varphi}^{(2)}=(\frac{iu_{-}}{\hat{\zeta}_{1}^{2}},\cdots, \frac{iu_{-}}{\hat{\zeta}_{4N_{1}+2N_{2}}^{2}})^{T},
\tilde{\varphi}^{(3)}=(\frac{2iu_{-}}{\hat{\zeta}_{1}^{3}}, \cdots, \frac{2iu_{-}}{\hat{\zeta}_{4N_{1}+2N_{2}}^{3}})^{T},\notag\\
\tilde{\gamma}=\left(\begin{array}{c}
     \tilde{\gamma}^{(1)}\\
  \tilde{\gamma}^{(2)}\\
  \tilde{\gamma}^{(3)}\\
\end{array}\right),\ \tilde{\gamma}^{(1)}=(\mu_{-11}(\zeta_{1}), \cdots, \mu_{-11}(\zeta_{4N_{1}+2N_{2}}))^{T},\tilde{\gamma}^{(2)}=(\mu'_{-11}(\zeta_{1}), \cdots, \mu'_{-11}(\zeta_{4N_{1}+2N_{2}}))^{T}, \notag\\
\tilde{\gamma}^{(3)}=(\mu''_{-11}(\zeta_{1}), \cdots, \mu''_{-11}(\zeta_{4N_{1}+2N_{2}}))^{T}, \mathcal{\tilde{H}}=\left(\begin{array}{ccc}
    \mathcal{\tilde{H}}^{(11)} &  \mathcal{\tilde{H}}^{(12)} &  \mathcal{\tilde{H}}^{(13)}\\
    \mathcal{\tilde{H}}^{(21)} &  \mathcal{\tilde{H}}^{(22)} &  \mathcal{\tilde{H}}^{(23)}\\
    \mathcal{\tilde{H}}^{(31)} &  \mathcal{\tilde{H}}^{(32)} &  \mathcal{\tilde{H}}^{(33)}\\
\end{array}\right),\label{97b}
\end{gather}
with $\mathcal{\tilde{H}}^{(im)}=\left(\mathcal{\tilde{H}}^{(im)}_{jn}\right)_{4N_{1}+2N_{2}\times 4N_{1}+2N_{2}}(i, m=
1, 2, 3)$ given by
\begin{gather}
\mathcal{\tilde{H}}^{(11)}_{jn}=C_{n}(\hat{\zeta}_{j})\left(\frac{1}{(\hat{\zeta}_{j}-\zeta_{n})^{2}}+\frac{D_{n}}{\hat{\zeta}_{j}-\zeta_{n}}+F_{n}\right)
+\frac{iu_{-}}{\hat{\zeta}_{j}}\delta_{j,n}, \
\mathcal{\tilde{H}}^{(12)}_{jn}=C_{n}(\hat{\zeta}_{j})\left(D_{n}+\frac{1}{\hat{\zeta}_{j}-\zeta_{n}}\right),\ \mathcal{\tilde{H}}^{(13)}_{jn}= \frac{1}{2}C_{n}(\hat{\zeta}_{j}),\notag\\
\mathcal{\tilde{H}}^{(21)}_{jn}=\frac{C_{n}(\hat{\zeta}_{j})}{\hat{\zeta}_{j}-\zeta_{n}}\left(\frac{3}{(\hat{\zeta}_{j}-\zeta_{n})^{2}}+\frac{2D_{n}}{\hat{\zeta}_{j}-\zeta_{n}}+F_{n}\right)
+\frac{iu_{-}}{\hat{\zeta}^{2}_{j}}\delta_{j,n}, \
\mathcal{\tilde{H}}^{(22)}_{jn}=\frac{C_{n}(\hat{\zeta}_{j})}{\hat{\zeta}_{j}-\zeta_{n}}\left(D_{n}+\frac{2}{\hat{\zeta}_{j}-\zeta_{n}}\right)-\frac{iu_{0}^{2}u_{-}}{\hat{\zeta}^{3}_{j}}\delta_{j,n},\notag\\
\mathcal{\tilde{H}}^{(23)}_{jn}=\frac{C_{n}(\hat{\zeta}_{j})}{2(\hat{\zeta}_{j}-\zeta_{n})},
\ \mathcal{\tilde{H}}^{(31)}_{jn}=\frac{2C_{n}(\hat{\zeta}_{j})}{(\hat{\zeta}_{j}-\zeta_{n})^{2}}\left(\frac{6}{(\hat{\zeta}_{j}-\zeta_{n})^{2}}+\frac{3D_{n}}{\hat{\zeta}_{j}-\zeta_{n}}+F_{n}\right)
+\frac{2iu_{-}}{\hat{\zeta}^{3}_{j}}\delta_{j,n},\notag\\ \mathcal{\tilde{H}}^{(32)}_{jn}=\frac{2C_{n}(\hat{\zeta}_{j})}{(\hat{\zeta}_{j}-\zeta_{n})^{2}}\left(D_{n}+\frac{3}{\hat{\zeta}_{j}-\zeta_{n}}\right)-\frac{4iu_{0}^{2}u_{-}}{\hat{\zeta}^{4}_{j}}\delta_{j,n},
\ \mathcal{\tilde{H}}^{(33)}_{jn}=\frac{C_{n}(\hat{\zeta}_{j})}{(\hat{\zeta}_{j}-\zeta_{n})^{2}}+\frac{iu_{0}^{4}u_{-}}{\hat{\zeta}^{5}_{j}}\delta_{j,n}.\label{98b}
\end{gather}
For the case of reflectionless potential, Eq.\eqref{38b} is denoted as
\begin{align}\label{99b}
u=u_{-}+i\tilde{\chi}^{T} \tilde{\gamma},
\end{align}
where
\begin{gather}
\tilde{\chi}=\left(\begin{array}{c}
     \tilde{\chi}^{(1)}\\
  \tilde{\chi}^{(2)}\\
  \tilde{\chi}^{(3)}\\
\end{array}\right),
\tilde{\chi}^{(1)}=(A[\zeta_{1}]e^{-2i\theta(\zeta_{1})}F_{1}, \cdots, A[\zeta_{4N_{1}+2N_{2}}]e^{-2i\theta(\zeta_{4N_{1}+2N_{2}})}F_{4N_{1}+2N_{2}})^{T},\notag\\
\tilde{\chi}^{(2)}=(A[\zeta_{1}]e^{-2i\theta(\zeta_{1})}D_{1}, \cdots, A[\zeta_{4N_{1}+2N_{2}}]e^{-2i\theta(\zeta_{4N_{1}+2N_{2}})}D_{4N_{1}+2N_{2}})^{T},\notag\\
\tilde{\chi}^{(3)}=(\frac{1}{2}A[\zeta_{1}]e^{-2i\theta(\zeta_{1})}, \cdots, \frac{1}{2}A[\zeta_{4N_{1}+2N_{2}}]e^{-2i\theta(\zeta_{4N_{1}+2N_{2}})})^{T}.\label{100b}
\end{gather}
Using Eqs. \eqref{96b}, the expression of the triple poles soliton solution can be derived finally.
\end{proof}

In particular, we exhibit the obtained solutions with $N_{1}=0, N_{2}=1$ and $N_{1}=1, N_{2}=0$, respectively. In the case of $N_{1}=0, N_{2}=1$, the one-triple poles soliton become dark-bright-dark solutions of the GI equation with NZBCs \eqref{49} in Fig.8 for the parameters  $u_{-}=1, A[\omega_{1}]=B[\omega_{1}]=C[\omega_{1}]=1, \omega_{1}=e^{\frac{\pi i}{6}}$ in
which the theta condition become $\mbox{arg}(\frac{u_{+}}{u_{-}})=2\pi$ such that $u_{+}=u_{-}$. On the other hand, when selecting $N_{1}=1, N_{2}=0$,  the one-triple poles soliton displays  breather-breather-breather in Fig. 9 for the parameters  $u_{-}=1, A[\omega_{1}]=B[\omega_{1}]=C[\omega_{1}]=1, \omega_{1}=2e^{\frac{\pi i}{6}}$ in which the theta condition is $\mbox{arg}(\frac{u_{+}}{u_{-}})=4\pi$ such that $u_{+}=u_{-}$. Besides, when $N_{1}=N_{2}=1$ i.e., a mixed discrete spectra, it turns into the three-breather-three-soliton solutions, meaning the interaction of three-breather and three-soliton.\\
{\rotatebox{0}{\includegraphics[width=5.0cm,height=5.0cm,angle=0]{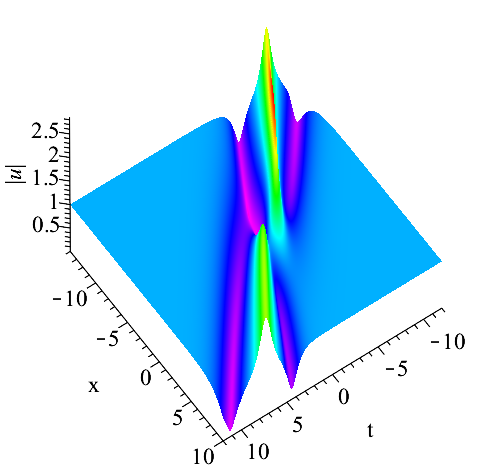}}}
~~~~\quad\qquad
{\rotatebox{0}{\includegraphics[width=5.0cm,height=5.0cm,angle=0]{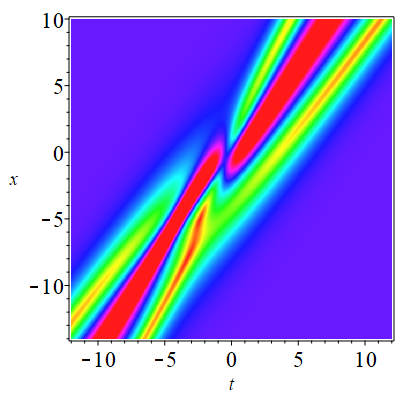}}}
~~~~\quad\qquad
{\rotatebox{0}{\includegraphics[width=5.0cm,height=5.0cm,angle=0]{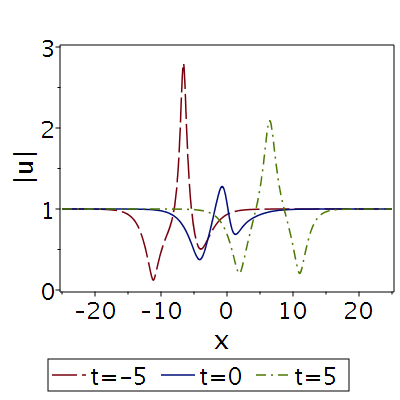}}}

$\qquad\qquad\textbf{(a)}\qquad\qquad\qquad\qquad\qquad\qquad\quad\qquad\qquad\quad 
\textbf{(b)}\qquad\qquad\qquad\qquad\qquad\qquad\qquad\qquad\quad\textbf{(c)}$\\
\noindent { \small \textbf{Figure 8.} (Color online) The one-triple pole soliton solution for Eq.\eqref{1} with NZBCs and $N_{1}=0, N_{2}=1$. The parameters are $u_{\pm}=1, A[\omega_{1}]=B[\omega_{1}]=C[\omega_{1}]=1, \omega_{1}=e^{\frac{\pi i}{6}}$. $\textbf{(a)}$ Three dimensional plot;
$\textbf{(b)}$ The density plot;
$\textbf{(c)}$ The wave propagation along the $x$-axis at $t=-5$(longdash), $t=0$(solid), $t=5$(dashdot).}\\
{\rotatebox{0}{\includegraphics[width=5.0cm,height=5.0cm,angle=0]{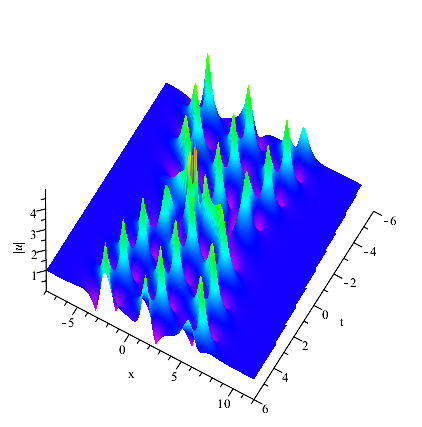}}}
~~~~\quad\qquad
{\rotatebox{0}{\includegraphics[width=5.0cm,height=5.0cm,angle=0]{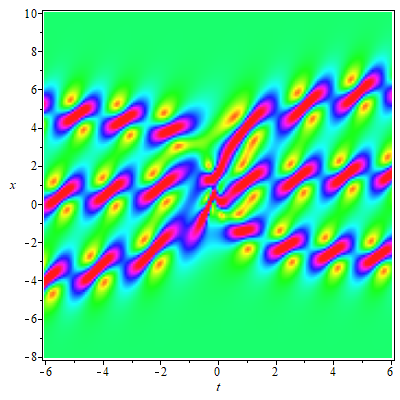}}}
~~~~\quad\qquad
{\rotatebox{0}{\includegraphics[width=5.0cm,height=5.0cm,angle=0]{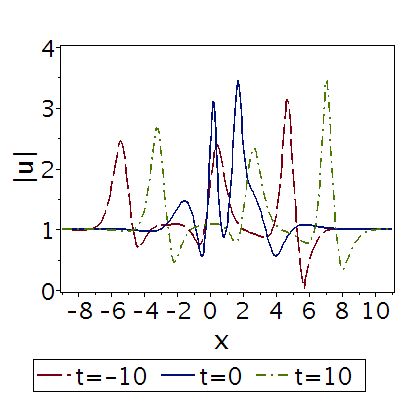}}}

$\qquad\qquad\textbf{(a)}\qquad\qquad\qquad\qquad\qquad\qquad\quad\qquad\qquad\quad 
\textbf{(b)}\qquad\qquad\qquad\qquad\qquad\qquad\qquad\qquad\quad\textbf{(c)}$\\
\noindent { \small \textbf{Figure 9.} (Color online) The one-triple pole soliton solution for Eq.\eqref{1} with NZBCs and $N_{1}=1, N_{2}=0$. The parameters are $u_{\pm}=1, A[\omega_{1}]=B[\omega_{1}]=C[\omega_{1}]=1, \omega_{1}=2e^{\frac{\pi i}{6}}$. $\textbf{(a)}$ Three dimensional plot;
$\textbf{(b)}$ The density plot;
$\textbf{(c)}$ The wave propagation along the $x$-axis at $t=-10$(longdash), $t=0$(solid), $t=10$(dashdot).}\\

\section{Conclusion}
In this paper, we have applied the RH method to discuss the GI type of derivative NLS equation with ZBCs and NZBCs. Through solving the RHP at the case of double and  triple poles, we have given out the $N$-double and $N$-triple poles soliton solutions under ZBCs and NZBCs.  The critical technique shown in this work is to eliminate the properties of singularities via subtracting the residue and the coefficient $L_{-2}$ from the original non-regular RHP when reflection coefficients have double poles. For the case of  triple poles, we have to subtract another the coefficient $L_{-3}$. Additionally, the asymptotic value of jump matrix is subtracted from the original non-regular RHP. Then the regular RHP can be displayed, which can be solved by Plemelj formula. Finally, the $N$-double and $N$-triple poles soliton solutions can be derived by using the solution of RHP to reconstruct potential function. Also, through choosing suitable parameters, the dynamic behaviors of one-double poles soliton, two-double poles soliton, one-triple poles soliton corresponding to ZBCs, one-double poles soliton, and one-triple poles soliton corresponding to NZBCs are analysed.  In the near future,  more works remain to be solved for other integrable systems via the technique shown in this paper.

\end{document}